\documentclass[11pt]{article}
\usepackage[a4paper,margin=1in]{geometry}
\usepackage{amsmath,amssymb}
\usepackage{graphicx}
\usepackage{hyperref}
\usepackage{cite}

 \usepackage{multirow}
\usepackage{tabularx, booktabs} 
\usepackage{tensor}
\usepackage{color} 
\usepackage{xcolor}

\usepackage{caption}  
\usepackage{subcaption} 

\usepackage{float}
\restylefloat{figure}

\pdfoutput=1

\def\be{\begin{equation}}
\def\ee{\end{equation}}
\def\ba{\begin{eqnarray}}
\def\ea{\end{eqnarray}}

\def\ga{\mathrel{\raise.3ex\hbox{$>$\kern-.75em\lower1ex\hbox{$\sim$}}}}
\def\la{\mathrel{\raise.3ex\hbox{$<$\kern-.75em\lower1ex\hbox{$\sim$}}}}

\newcommand{\fr}[2]{\frac{#1}{#2}}

\newcommand{\oo}{\omega_{0}}
\newcommand{\oa}{\omega_{a}}

\newcommand{\rhom}{\rho_{\rm{m}}}
\newcommand{\rhode}{\rho_{\rm{DE}}}
\newcommand{\Ode}{\Omega_{\rm{DE} 0}}
\newcommand{\Omo}{\Omega_{\rm{m} 0}}
\newcommand{\Omz}{\Omega_{\rm{m}}(z)}

\newcommand{\EG}{E_{\rm{G}}}

\newcommand{\DL}{D_{\rm{L}}}

\title{Probing Time-Varying Dark Energy with DESI: The Crucial Role of Precision Matter Density ($\Omo$) Measurements}
\author{Seokcheon Lee\\
\small Department of Physics, Institute of Basic Science, Sungkyunkwan University, Suwon 16419, Korea\\
\small \texttt{skylee@skku.edu}}
\date{\today}

\begin{document}

\maketitle

\begin{abstract}

Accurate measurements of fundamental cosmological parameters, especially the Hubble constant ($H_0$) and present-day matter density ($\Omo$), are crucial for constraining dark energy (DE) evolution. We analyze the sensitivities of cosmological observables ($H(z)$, $D_L(z)$, $\text{E}_{\text{G}}$) to $\Omo$, $\oo$, and $\oa$ under different parametrizations. Our results show observables are far more sensitive to $\Omo$ than to DE equation of state parameters (e.g., at $z \sim 0.5$, $H(z)$'s $\Omo$ sensitivity is $\sim 0.7$ vs. $\oa$'s $\sim 0.04$). This hierarchy mandates high-precision $\Omo$ measurements to accurately constrain time-varying DE.  We also find DE parameter sensitivity highly depends on parametrization; the standard CPL form shows low sensitivity to $\oa$, but $\omega(z) = \oo + \oa \ln(1+z)$ significantly enhances it. Our analysis of DESI DR1/DR2 data confirms these theoretical limits: standalone DESI data primarily provides only upper limits for $\oa$, underscoring insufficient constraining power for a definitive time-varying DE detection. While combined datasets offer tighter constraints, interpretation requires caution due to parametrization influence.  We further confirm this point using simulated Supernovae MCMC data.In conclusion, improving $\Omo$ precision and adopting optimized parametrizations are imperative for future surveys like DESI to fully probe dark energy's nature.

\end{abstract}

\tableofcontents
\newpage

\section{Introduction}
\setcounter{equation}{0}

Recent precise astrophysical and cosmological measurements, utilizing techniques such as baryon acoustic oscillations (BAO) \cite{SDSS:2005xqv,2dFGRS:2005yhx,Blake:2011en,BOSS:2012dmf,eBOSS:2015jyv,BOSS:2016wmc,DES:2024pwq,DESI:2024mwx,DESI:2025zgx}, cosmic microwave background (CMB) anisotropies \cite{Planck:2018vyg,Planck:2018jri,Planck:2019kim}, Type Ia supernovae (SNe Ia) \cite{Scolnic:2021amr,Brout:2022vxf,DES:2024jxu}, weak lensing \cite{Heymans:2020gsg,DES:2021bvc,Dalal:2023olq,Sugiyama:2023fzm}, and galaxy velocities  \cite{Bhattacharya:2010cf,Agrawal:2019yed,Zhang:2024pyf}, consistently support a spatially flat universe dominated by dark energy (DE) in the form of a cosmological constant, known as the $\Lambda$CDM model. The curvature density parameter $\Omega_k$ is constrained to be very close to zero, a prediction also strongly supported by inflationary cosmology, which addresses the flatness problem. In this work, we consider a flat ($\Omega_k = 0$) $\omega$CDM universe as our fiducial model, where $\omega$ denotes a constant DE equation of state (EOS) that reduces to the $\Lambda$CDM model when $\omega = -1$. For this framework, the Hubble parameter $H(z)$, angular diameter distance $D_A(z)$, luminosity distance $D_L(z)$, and the linear growth factor of matter density perturbations $\delta_m(z)$ possess analytical solutions depending solely on the present-day matter density $\Omo$ and $\omega$. To explore the possibility of a time-varying EOS for DE, we adopt the widely used Chevallier-Polarski-Linder (CPL) parametrization, given by $\omega(z) = \omega_{0} + \omega_{a} \frac{z}{1+z}$, characterized by its present-day value ($\omega_{0}$) and its evolutionary rate ($\omega_{a}$) \cite{Chevallier:2000qy,Linder:2002et}. 

Recent studies have emphasized that cosmological inferences on DE can be significantly affected by the choice of parametrization itself. In particular, alternative parametrizations of the Hubble parameter or the dark-energy equation of state have been explored in modified gravity and phenomenological frameworks, highlighting nontrivial degeneracies and model-dependent interpretations~\cite{Myrzakulov:2023qjo,Myrzakulov:2023sir,Shekh:2023vtq}. These works reinforce the need for a careful assessment of how specific parametrization choices influence the inferred dynamics of DE, especially when extending beyond a constant equation of state.

One crucial observational avenue for constraining cosmological parameters is the direct measurement of $H(z)$, for which several independent techniques have been developed. These include the analysis of radial BAO \cite{Seo:2003pu,Gaztanaga:2008xz,Moresco:2012by,BOSS:2016wmc} and the differential ages of passively evolving galaxies (cosmic chronometers) \cite{Jimenez:2003iv,Simon:2004tf,Stern:2009ep,Borghi:2021rft,Jiao:2022aep,Moresco:2024wmr}. The latter is suggested to offer potentially higher parameter constraining power than current SNe Ia datasets if more independent measurements across a wider redshift range ($0 \leq z \leq 2$) can be obtained \cite{Ma:2010mr}, though limitations due to lensing covariance in small areas might exist \cite{Cooray:2005yr}. Additionally, $H(z)$ can be inferred from the linear differentiation of measured comoving distances from SNe Ia in uncorrelated redshift bins \cite{Wang:2005yaa}, and from the power spectrum of peculiar velocities measured through the kinematic Sunyaev-Zel'dovich (kSZ) effect \cite{Hernandez-Monteagudo:2005xtx,Bhattacharya:2007sk,Hand:2012ui,Tanimura:2020une,DES:2023mug}. While kSZ is less sensitive to systematic redshift errors compared to luminosity distance or weak lensing techniques, the reconstruction of the large-scale peculiar velocity field remains susceptible to observational biases \cite{Lavaux:2007zw}.  In addition to these approaches, recent works have investigated reconstructions of the Hubble parameter directly from observational data, aiming to reduce model dependence and assess systematic effects in late-time cosmology~\cite{Koussour:2024kxd,Alfedeel:2024ktc}. Such analyses provide an important complementary perspective on the robustness of $H(z)$ measurements and their implications for dynamical dark energy (DDE).

Alongside direct $H(z)$ measurements, angular diameter distance $D_A(z)$ and luminosity distance $D_L(z)$ are crucial geometrical probes for constraining cosmological parameters and the nature of DE. $D_A(z)$ can be determined using standard rulers like transverse BAO \cite{SDSS:2005xqv,eBOSS:2015jyv,SDSS:2009ocz,BOSS:2013rlg} and through strong gravitational lensing systems \cite{Chantry:2010ym,H0LiCOW:2019pvv}. Conversely, $D_L(z)$ is primarily measured using standard candles, most notably SNe Ia \cite{Brout:2022vxf,SupernovaSearchTeam:1998fmf,SupernovaCosmologyProject:1998vns,Pan-STARRS1:2017jku}, which played a pivotal role in the discovery of cosmic acceleration and continue to refine the distance-redshift relation, providing insights into DE properties. While the complementarity between $H(z)$ measurements (a direct probe of the expansion rate) and distance measurements ($D_A(z)$ and $D_L(z)$, which are integrals of the expansion history) is vital for breaking parameter degeneracies, it is important to note that for a time-varying DE EOS, $D_A(z)$ and $D_L(z)$ generally exhibit weaker constraining power compared to $H(z)$. This is because $D_L(z)$ and $D_A(z)$ and $H(z)$ depend on a double and single integral of the EOS with respect to redshift, respectively, effectively smoothing out the sensitivity to rapid variations in the DE properties over cosmic time. Therefore, while essential, distance measurements might be less sensitive to the detailed temporal evolution of DE compared to direct measurements of the Hubble parameter at various redshifts.

To further break degeneracies inherent in geometric tests of cosmology, the linear growth of large-scale structures provides crucial complementary information. As a test of General Relativity (GR) on cosmological scales, the $\EG$ parameter, defined as the ratio of the Laplacian of the Newtonian potential to the peculiar velocity divergence \cite{Zhang:2007nk}, offers a promising avenue. While current observational uncertainties preclude definitive conclusions, $\EG$ holds significant potential for probing the validity of GR \cite{Reyes:2010tr}.

To quantify the sensitivity of various observables to cosmological parameters, we employ their logarithmic derivatives with respect to the parameters. As shown in our analysis, for all three observables ($H(z)$, $D_L(z)$, and $\EG$), the sensitivity to $\Omo$ is consistently much higher than to the DE EOS parameters $\oo$ and especially $\oa$. For $H(z)$ and $\EG$, the sensitivities to $\oa$ typically peak at intermediate redshifts ($z \sim 0.5 - 1.0$) and are generally quite low, making $\oa$ challenging to constrain. In contrast, for $D_L(z)$, the absolute sensitivity to DE parameters often increases with redshift. This inherent difference in sensitivities across parameters and observables underscores the importance of precise $\Omo$ measurements to break degeneracies and highlights the complementary nature of different cosmological probes in constraining DE properties.

In this work, we perform a detailed sensitivity analysis of key cosmological observables to fundamental cosmological parameters, particularly focusing on the role of the present-day matter density ($\Omo$) in constraining the time-varying DE EOS ($\oo, \oa$). The present work is intended to provide the conceptual and analytic foundation for understanding prior-induced biases in DDE inference, which we later quantified explicitly using DESI DR2 BAO data in Ref.~\cite{Lee:2025kbn}, providing a complementary numerical validation of the mechanisms identified here.We compare the sensitivities of geometrical probes, including the Hubble parameter ($H(z)$) and luminosity distance ($D_L(z)$), with those of the growth-sensitive $\EG$ parameter. One can consider general DE EOS parametrization $\omega(z) = \oo + \oa f(z)$. It has been known that if $f(z)$ increases rapidly with redshift $z$, then the error in $\oa$ is smaller; conversely, if $f(z)$ increases slowly, the error in $\oa$ is larger. Therefore, the error in $\oa$ is known to be highly sensitive to the form of $f(z)$ \cite{Lee:2011ec,Colgain:2021pmf}. Furthermore, we investigate how the choice of DE parametrization, comparing the canonical CPL model with an alternative $\omega(z) = \oo + \oa \ln(1+z)$ form \cite{Efstathiou:1999tm}, impacts these sensitivities and the overall constraining power. 

While Principal Component Analysis (PCA) is often employed to overcome the limitations of specific DE EOS parametrizations, one must exercise caution when using this method due to inherent drawbacks. For instance, PCA faces limitations such as a restricted number of independent free parameters (e.g., $j-1$ for $j$ redshift bins) and its potential inability to accurately reproduce the true behavior of $\omega(z)$ if $\omega(z)$ varies non-linearly or with a small but non-negligible gradient, which can lead to confusion with a constant $\omega$ value \cite{Lee:2010nb}. Our analysis aims to identify optimal strategies for leveraging upcoming large-scale structure surveys like DESI to robustly probe the nature of DE.

The paper is organized as follows: In Section~\ref{sec:COTS}, we present the cosmological observables and their logarithmic derivatives with respect to the cosmological parameters. Section~\ref{sec:MAA} describes our methodology, including the DE parametrizations and analysis approach. The implications for time-varying DE EOS obtained from DESI are discussed in Section~\ref{sec:sensitivities}. Finally, we conclude with a summary of our findings and outline future prospects in Section~\ref{sec:Con}.

\section{Cosmological Observables and Their Sensitivities to Cosmological Parameters}
\label{sec:COTS}
\setcounter{equation}{0}

In this section, we present an analytic sensitivity analysis of cosmological observables with respect to the parameters $(\Omo, \oo, \oa)$. We focus on comparing the relative sensitivity hierarchies and degeneracy structures of geometric and structure-growth observables.

\subsection{Hubble Parameter ($H(z)$)}
\label{subsec:Hz}

Compared to cosmological distances such as the luminosity distance ($D_L(z)$) or the angular diameter distance ($D_A(z)$), the Hubble parameter $H(z)$ generally exhibits less degeneracy in constraining the DE EOS.  This advantage stems from its direct dependence on redshift, requiring one fewer integration step compared to distance measures~\cite{Astier:2000as,Huterer:2000mj,Maor:2001ku}. We emphasize that no DESI BAO likelihood or covariance matrix is used in the present analysis; DESI results are cited solely for contextual motivation. Assuming a spatially flat universe and neglecting the contribution of relativistic particles, the Hubble parameter within the CPL parametrization of the DE EOS ($\omega(z)=\oo+\oa z/(1+z)$) is given by
\be
H^2(z) = H_0^2 \left[ \Omo (1+z)^3 + (1-\Omo) (1+z)^{3(1+\oo+\oa)} e^{-3 \oa \frac{z}{1+z}} \right] . \label{Hz}
\ee
For later convenience and for direct comparison with the analytic derivatives presented in Appendix~\ref{app:dlnH}, we express $H(z)$ in this standard additive form.

\begin{figure}
\centering
\vspace{1.5cm}
\begin{tabular}{ccc}
\includegraphics[width=0.3\linewidth]{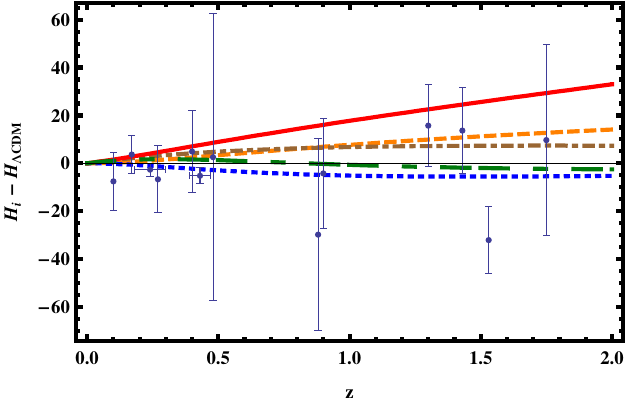} &
\includegraphics[width=0.3\linewidth]{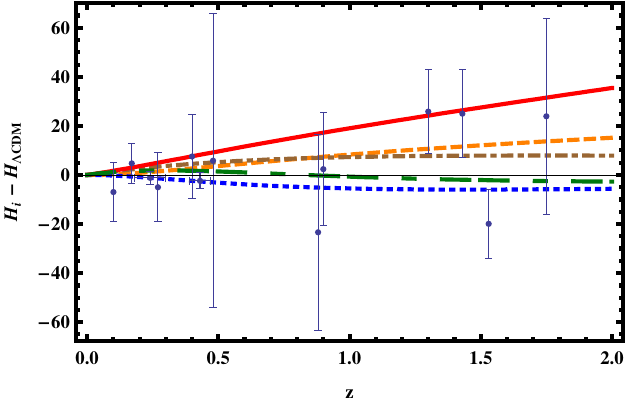} &
\includegraphics[width=0.3\linewidth]{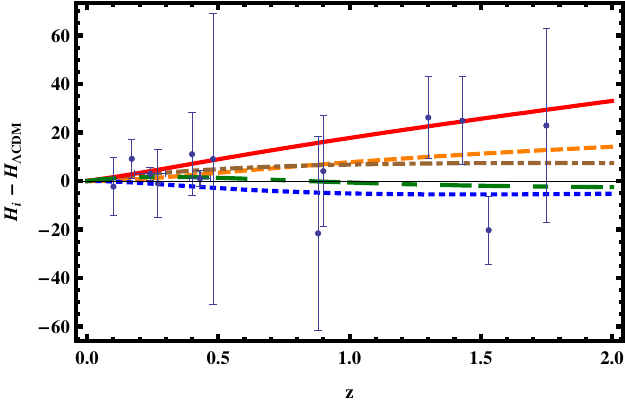} \\
\end{tabular}
\vspace{-0.5cm}
\caption{Relative deviations of $H$ from the $\Lambda$CDM prediction for different values of cosmological parameters
($H_0,\Omo$)=($73,0.3$), ($73,0.25$), and ($68,0.3$) from left to right. Observational data are taken from~\cite{Jimenez:2003iv,Stern:2009ep,Gaztanaga:2008xz}.  The observational data are shown for reference only. No likelihood fit or parameter inference is performed using these data in this figure.}
\label{fig1}
\end{figure}

Figure~\ref{fig1} illustrates the relative differences in $H(z)$ between various DE models and the fiducial $\Lambda$CDM model, shown for different values of $(H_0,\Omo)$.  The observational data points shown in the figure are included for reference only; no likelihood fit or statistical inference is performed.

Each panel presents models with distinct $(\oo,\oa)$ values, represented by specific colors and line styles: (red, solid, $-0.8,+0.8$), (orange, dashed, $-1.0,0.8$), (brown, dash-dotted, $-0.8,0$), (green, long-dashed, $-0.8,-0.8$), and (blue, dotted, $-1.0,-0.8$). These trends can be understood by analyzing Eq.~\eqref{Hz}, with $\Lambda$CDM serving as the reference case
($\oo=-1$, $\oa=0$), for which $H_{\Lambda{\rm CDM}}(z)= H_0\sqrt{\Omo(1+z)^3+(1-\Omo)}$.

\begin{itemize}
\item \textbf{For $\oa>0$ (red solid, orange dashed):}
\begin{itemize}
\item The exponential factor $e^{-3\oa z/(1+z)}$ decreases with increasing $z$.
\item The power-law term $(1+z)^{3(1+\oo+\oa)}$ can partially compensate for this suppression.
\item The combined effect may lead to a more rapid growth of $H(z)$ relative to $\Lambda$CDM for certain parameter choices.
\end{itemize}

\item \textbf{For $\oa=0$ (brown dash-dotted):}
\begin{itemize}
\item Equation~\eqref{Hz} reduces to $H^2(z)=H_0^2\left[\Omo(1+z)^3+(1-\Omo)(1+z)^{3(1+\oo)}\right]$.
\item For $\oo>-1$, the DE contribution decreases with redshift, leading to a smaller $H(z)$ at high $z$ compared to $\Lambda$CDM.
\end{itemize}

\item \textbf{For $\oa<0$ (green long-dashed, blue dotted):}
\begin{itemize}
\item The DE density decreases rapidly with redshift, driving the universe toward matter domination.
\item As a result, $E^2(z)\simeq \Omo(1+z)^3$ at high $z$, and the sensitivity of $H(z)$ to the time-varying DE component becomes strongly suppressed.
\end{itemize}
\end{itemize}

The plots also demonstrate that variations in $(H_0,\Omo)$ can mimic changes in the DE EOS over the observed redshift range. While marginalizing over $H_0$ with a Gaussian prior has a negligible impact on the relative sensitivity structure captured by our $\chi^2_H$ diagnostics (see subsection~\ref{subsec:PCA}), Fig.~\ref{fig1} indicates that current $H(z)$ data alone cannot definitively rule out the DE models considered here.

Furthermore, the model dependence of $H(z)$ becomes more pronounced at higher redshifts ($z\gtrsim1$). The redshift dependence of the sensitivities to $\oo$ and $\oa$ is governed by the analytic derivatives of $\ln H$ presented in Appendix~\ref{app:dlnH}.

In particular, for negative $\oa$, the rapid approach to matter domination causes the fractional sensitivity of $H(z)$ to $\oa$ to decrease sharply at high redshift, making it difficult to constrain time-varying DE using $H(z)$ data alone.

To further quantify these effects, we consider the logarithmic derivatives
\be
\fr{dH}{H} = \fr{\partial \ln H}{\partial \Omo} d\Omo + \fr{\partial \ln H}{\partial \oo} d\oo + \fr{\partial \ln H}{\partial \oa} d\oa ,
\label{dH}
\ee
where the coefficients
$\partial \ln H / \partial p_j$ with $p_j=\{\Omo,\oo,\oa\}$ are evaluated using the exact analytic expressions given in Appendix~\ref{app:dlnH} and form the basis of the Fisher-matrix analysis discussed in Sec.~\ref{sec:MAA}.

\begin{figure}
\centering
\vspace{1.5cm}
\begin{tabular}{ccc}
\includegraphics[width=0.3\linewidth]{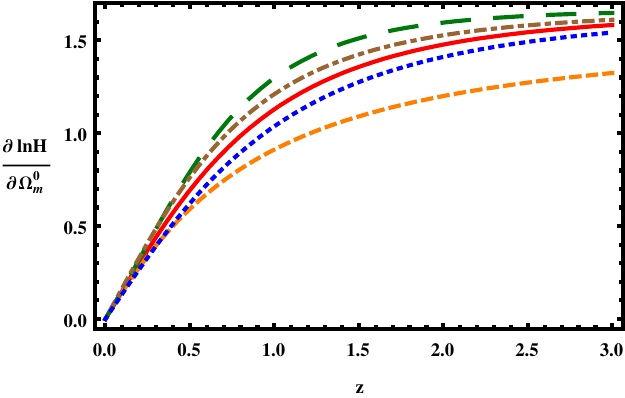} &
\includegraphics[width=0.3\linewidth]{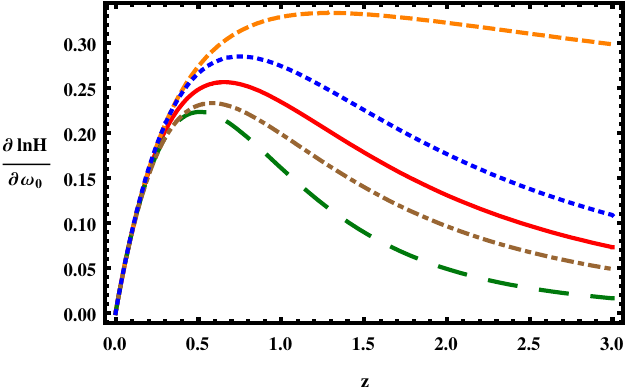} &
\includegraphics[width=0.3\linewidth]{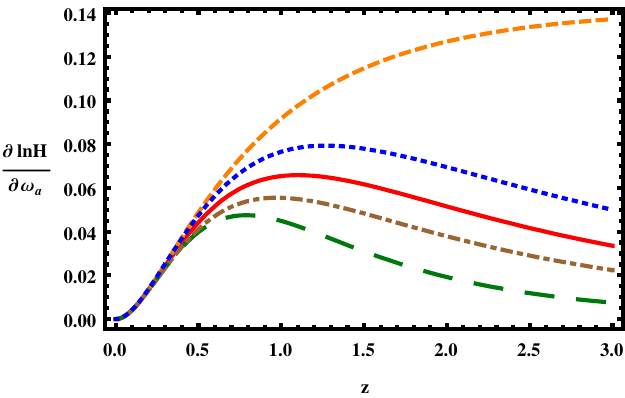} \\
\end{tabular}
\vspace{-0.5cm}
\caption{ Logarithmic derivatives of the Hubble parameter with respect to (w.r.t) $\Omo$ (left), $\oo$ (middle), and $\oa$ (right) as a function of redshift for different DE models: ($\oo$\,,$\oa$) = ($-1.0$, $-0.8$) (green, long-dashed), (-1.1, 0.0) (brown, dot-dashed), (-1.0, 0.0) (red, solid), (-0.9, 0.0) (blue, dotted), (-1.0, +0.8) (orange, dashed). } \label{fig2}
\end{figure}

Figure~\ref{fig2} displays these logarithmic derivatives for various DE models. The left panel demonstrates that $\partial \ln H / \partial \Omo$ has the largest magnitude among the three coefficients, indicating that the Hubble parameter is most sensitive to the present-day matter density. Around intermediate redshifts ($z\sim0.5$), the magnitude of $\partial \ln H / \partial \Omo$ is substantially larger than those of $\partial \ln H / \partial \oo$ and $\partial \ln H / \partial \oa$ for the fiducial $\Lambda$CDM model, revealing a clear hierarchy in parameter sensitivities. This hierarchy highlights the significantly lower sensitivity of $H(z)$ to the DE EOS parameters compared to the matter density, and underscores the necessity of accurate $\Omo$ measurements for effectively constraining the DE EOS. For rapidly evolving DE models with positive $\oa$ (orange dashed line), the sensitivity to $\Omo$ is comparatively lower due to the increased contribution of DE. Figure~\ref{fig2} illustrates relative parameter sensitivities and degeneracy directions; it does not represent forecasted constraints from a specific experiment.

The middle and right panels of Fig.~\ref{fig2} reveal the redshift dependence of $\partial \ln H / \partial \oo$ and $\partial \ln H / \partial \oa$, respectively. For positive values of $\oa$, a model-dependent redshift $z_\ast$ exists at which these sensitivities reach their maximum, indicating an optimal redshift range for constraining the corresponding DE parameters.

However, Fig.~\ref{fig2} also underscores the challenge in precisely constraining $\oa$, as the magnitude of $\vec{g}(z,\oa \equiv\partial\ln H/\partial\oa$ remains relatively small compared to $\vec{g}(z,\oo)$ for most considered models, particularly for slowly varying or negative $\oa$. While this conclusion is model-dependent, alternative parametrizations of $\omega(z)$ can potentially enhance the sensitivity to $\oa$, with stringent constraints on $\omega(z)$ typically occurring around $z\sim0.5$, consistent with previous findings~\cite{Alam:2003fg,Wu:2007tz}.

The sensitivity curves shown in Fig.~\ref{fig2} can be understood as follows:
\begin{itemize}
    \item \textbf{Sensitivity to $\Omo$ (Left Panel):} \\
    All curves generally increase with redshift $z$. The long-dashed (green) line ($\oo=-1.0,\oa=-0.8$) exhibits the highest sensitivity across the entire redshift range, indicating that in this model a small change in $\Omo$ leads to a relatively larger change in $H(z)$. The dot-dashed (brown) line ($\oo=-1.1,\oa=0.0$) shows the second-highest sensitivity. The solid (red) line ($\oo=-1.0,\oa=0.0$, corresponding to $\Lambda$CDM) displays a moderate increase, while the dotted (blue) line ($\oo=-0.9,\oa=0.0$) shows a slightly lower sensitivity. Conversely, the dashed (orange) line ($\oo=-1.0,\oa=+0.8$) exhibits the lowest sensitivity, which can be attributed to the larger contribution of DE at higher redshifts, reducing the relative influence of matter density.
    \item \textbf{Sensitivity to $\oo$ (Middle Panel):} \\
    The sensitivity of $H(z)$ to $\oo$ initially increases with redshift, reaches a peak at intermediate redshifts, and then decreases. This behavior indicates that the impact of the present-day DE EOS is most significant at intermediate epochs. For rapidly evolving DE models with positive $\oa$ (orange dashed line), the peak sensitivity occurs at relatively higher redshift, whereas for constant or slowly evolving models ($\oa\simeq0$) the peak shifts to slightly lower redshifts. Models with more negative $\oo$ (brown dot-dashed line) show a reduced overall sensitivity. In contrast, for models with negative $\oa$ (green long-dashed line), the sensitivity remains low across all redshifts as the universe quickly approaches matter domination.
    \item \textbf{Sensitivity to $\oa$ (Right Panel):} \\
    The sensitivity to $\oa$ generally increases with redshift for most models, reflecting the cumulative effect of a time-varying DE component on the expansion history. For models with positive $\oa$ (orange dashed line), this sensitivity is highest across a broad redshift range. In contrast, for constant or mildly evolving DE models ($\oa\simeq0$), the sensitivity peaks at intermediate redshifts and decreases thereafter. Models with negative $\oa$ (green long-dashed line) exhibit a much weaker sensitivity at low redshift, which increases toward higher redshift but remains subdominant. These trends can be directly understood from the analytic expressions for the logarithmic derivatives of $H(z)$ given in Appendix~\ref{app:dlnH}, which contain terms proportional to $\ln(1+z)$ and are strongly suppressed once the expansion becomes matter dominated.
\end{itemize}

\subsection{Luminosity Distance ($D_L(z)$)}
\label{subsec:DL}

\begin{figure}
\centering
\vspace{1.5cm}
\begin{tabular}{ccc}
\includegraphics[width=0.3\linewidth]{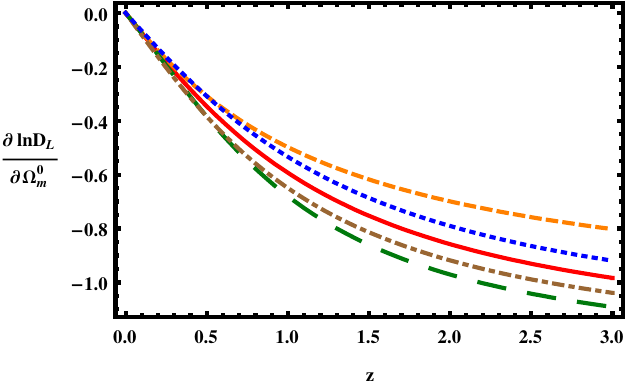} &
\includegraphics[width=0.3\linewidth]{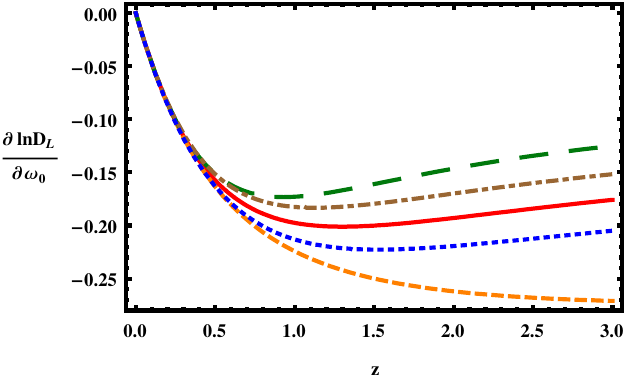} &
\includegraphics[width=0.3\linewidth]{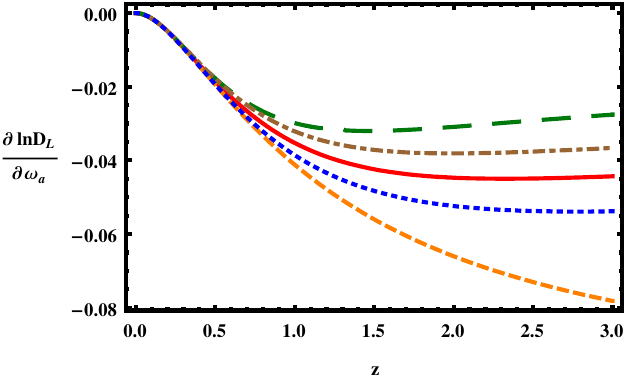} \\
\end{tabular}
\vspace{-0.5cm}
\caption{ Logarithmic derivatives of the luminosity distance w.r.t $\Omo$ (left), $\oo$ (middle), and $\oa$ (right) as a function of redshift for the different DE models. The representation of each curve corresponds to the same DE models used in Figure~\ref{fig2}.} \label{fig3}
\end{figure}

We now extend our sensitivity analysis to the luminosity distance, $\DL(z) = (1+z) \int_{0}^{z} \fr{dz'}{H(z')}$, a key geometric quantity derived from SNe Ia observations. While the degeneracy of $\Omo$ in $\DL$ has been extensively studied elsewhere \cite{Astier:2000as,Huterer:2000mj,Maor:2001ku,Colgain:2022tql}, our focus here is on the differential error in $\DL(z)$, given by 
\begin{equation}
\fr{d\DL}{\DL} = \fr{\partial \ln \DL}{\partial \Omo} d \Omo + \fr{\partial \ln \DL}{\partial \oo} d \oo + \fr{\partial \ln \DL}{\partial \oa} d \oa \label{dDL} \,.
\end{equation}
The logarithmic derivatives, $\partial \ln \DL / \partial \Omo$, $\partial \ln \DL / \partial \oo$, and $\partial \ln \DL / \partial \oa$, are plotted against redshift $z$ in Figure \ref{fig3} (left, middle, and right panels, respectively). The curves in Figure \ref{fig3} correspond to the same DE models defined in Figure \ref{fig2}: ($-1.0, 0$) (red solid), ($-1.0, 0.8$) (orange dashed), ($-1.0, -0.8$) (green long-dashed), ($-1.1, 0$) (brwon dot-dashed), and ($-0.9, 0$) (blue dotted). Consistent with $H(z)$, the luminosity distance is most strongly sensitive to $\Omo$. However, compared to $H(z)$, the sensitivities on DE EOS parameters are generally weaker in $\DL(z)$ due to its double integral dependence on cosmological parameters. Similar to $H(z)$, the rapidly time varying $\omega$ model with positive $\oa$ (orange dashed line) shows the least sensitivity to $\Omo$ among the considered models, attributable to its larger DE contribution.  Moreover, the existence of a sharp peak for the most sensitive redshift for DE parameters, as seen in $H(z)$, is less pronounced in $\DL(z)$. Since the angular diameter distance, $D_A = \DL / (1+z)^2$, is directly proportional to $\DL(z)$, a separate investigation of its sensitivity to these parameters is redundant. Thus, $\DL$ can be used to provide complementary constraints on $\oo$ and $\oa$, alongside $H$. Although the DE density is subdominant at high redshift for positive $\oa$, its evolving contribution modifies the integrated expansion history, leading to a non-negligible cumulative effect on $D_L(z)$.

\begin{itemize}
    \item \textbf{Sensitivity to $\Omo$ (Left Panel):} \\
The sensitivity of the luminosity distance to the present-day matter density, $\Omo$, as depicted in the left panel of Figure~\ref{fig3}, shows that $\frac{\partial \ln D_L}{\partial \Omo}$ is negative and its absolute value generally increases with redshift for all considered DE models. This indicates that the luminosity distance becomes more sensitive to changes in $\Omo$ at higher redshifts. Physically, a higher matter density leads to a slower expansion rate at earlier times, resulting in smaller luminosity distances to high-redshift objects. The different DE models exhibit slightly varying degrees of this sensitivity. The long-dashed (green) line ($\oo = -1.0, \oa = -0.8$) shows the largest absolute sensitivity to $\Omo$ at high redshifts, suggesting that for DE models where its density was less dominant in the past (more negative $\oa$ implies DE density decreases with redshift, making matter relatively more dominant at higher $z$), the impact of matter density on the observed $D_L$ is most pronounced at early epochs. Conversely, the dashed (orange) line ($\oo = -1.0, \oa = +0.8$) displays the weakest absolute sensitivity at high redshifts, implying that for models where DE density increases more significantly towards the past (positive $\oa$ means DE density was smaller in the past, making the universe more matter-dominated; however, the influence of DE growth is also present), the influence of matter density on the luminosity distance is somewhat diminished at early times compared to other models. The solid (red) ($\Lambda$CDM), dotted (blue), and dot-dashed (brown) lines show intermediate sensitivities, with the $\Lambda$CDM model falling within this range. Overall, the strong dependence of the luminosity distance on the matter density parameter across all redshifts confirms that supernovae observations are a crucial tool for constraining $\Omo$.

	\item \textbf{Sensitivity to $\oo$ (Middle Panel):} \\
The sensitivity of the $D_L$ to the present-day value of the DE EOS, $\oo$, as shown in the middle panel of Figure~\ref{fig3}, reveals that $\frac{\partial \ln D_L}{\partial \oo}$ is also negative for all considered models and exhibits a non-monotonic behavior with redshift. At low redshifts, the sensitivity is close to zero, indicating that the current value of $\oo$ has a minimal impact on the luminosity distance to nearby objects. As redshift increases, the absolute sensitivity grows, reaching a maximum (most negative value) around $z \sim 0.7$, suggesting that $\oo$ has the strongest influence on the luminosity distance to objects at these intermediate redshifts. Beyond this redshift range, the absolute sensitivity tends to decrease slightly for some models. Physically, $\oo$ governs the current energy density and pressure of DE, which in turn affects the expansion history and thus the integrated luminosity distance. The delayed impact of $\oo$ on $D_L$ compared to $H(z)$ is expected due to the integral nature of the luminosity distance. The dashed (orange) line ($\oo = -1.0, \oa = +0.8$) shows the largest absolute sensitivity to $\oo$ at intermediate redshifts, implying that for rapidly evolving DE models, the current value of the EOS has a significant impact on the observed distances to objects in this redshift range. The long-dashed (green) line ($\oo = -1.0, \oa = -0.8$) displays the smallest absolute sensitivity at higher redshifts, suggesting a weaker influence of $\oo$ on $D_L$ for models where DE density was stronger in the past (more negative $\oa$ means DE density decreases less steeply or even increases towards the past, making DE more dominant at high $z$ and thus reducing the relative impact of $\oo$). The $\Lambda$CDM (red solid), dotted, and dot-dashed lines show intermediate sensitivities. The non-zero sensitivity across a significant redshift range indicates that supernovae observations can provide constraints on the present-day value of the DE EOS.

	 \item \textbf{Sensitivity to $\oa$ (Right Panel):} \\
The sensitivity of the luminosity distance to $\oa$ as shown in the right panel of Figure~\ref{fig3}. The luminosity distance becomes more sensitive to $\oa$ at higher redshifts. Physically, $\oa$ governs how the DE density and pressure evolve over cosmic time, and its impact on the integrated $D_L$ becomes more pronounced when considering light from more distant objects that has traveled through a longer and differently expanding universe. The dashed (orange) line ($\oo = -1.0, \oa = +0.8$) exhibits the largest absolute sensitivity to $\oa$ at high redshifts, suggesting that for rapidly evolving DE models where the EOS changes significantly with time, the effect on the observed luminosity distance to distant supernovae is most substantial. Conversely, the long-dashed (green) line ($\oo = -1.0, \oa = -0.8$) shows the smallest absolute sensitivity at higher redshifts, implying a weaker influence of $\oa$ on $D_L$ for models where the DE density evolution is such that its impact on the overall expansion history is less sensitive to $\oa$. The $\Lambda$CDM (red solid), dotted, and dot-dashed lines display intermediate sensitivities. The increasing absolute sensitivity with redshift for most models suggests that high-redshift supernovae observations are particularly valuable for constraining the time evolution of the DE EOS. However, it is important to note that the overall magnitude of the sensitivity to $\oa$ is considerably smaller compared to that for $\Omo$ and $\oo$, reinforcing the challenge of precisely constraining $\oa$ solely with luminosity distance measurements.
\end{itemize}

\subsection{The $\EG$ Parameter: A Probe of Gravity and Growth of Structure}
\label{subsec:EG}

In addition to the geometric tests, which often suffer from parameter degeneracies, the linear growth of large-scale structure provides a crucial avenue for breaking these degeneracies. As a powerful tool to test GR on cosmological scales, the $\EG$ statistic is defined as the ratio of the Laplacian of the Newtonian potentials to the peculiar velocity divergence ($\EG \equiv \fr{\nabla^2 \Phi}{\theta H_0^2 a^{-1}}$) \cite{Zhang:2007nk}. Within linear perturbation theory, this quantity can be expressed as $\EG(z) = \fr{\Omo}{f(z)}$, where $f(z) \equiv \fr{d \ln \delta}{d \ln a} = - (1+z) \fr{d \ln \delta}{dz}$ is the linear growth rate. 

\begin{figure}
\begin{tabular}{ccc}
\includegraphics[width=0.3\linewidth]{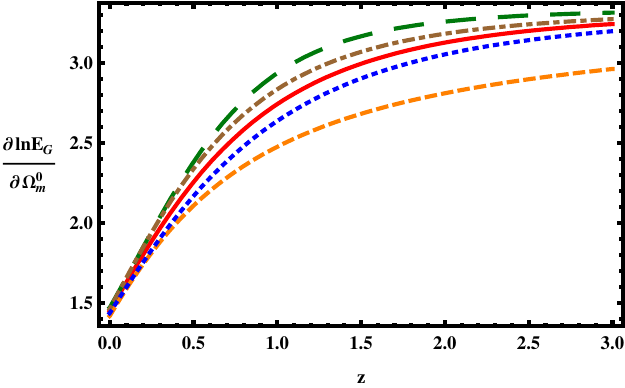}  &
\includegraphics[width=0.3\linewidth]{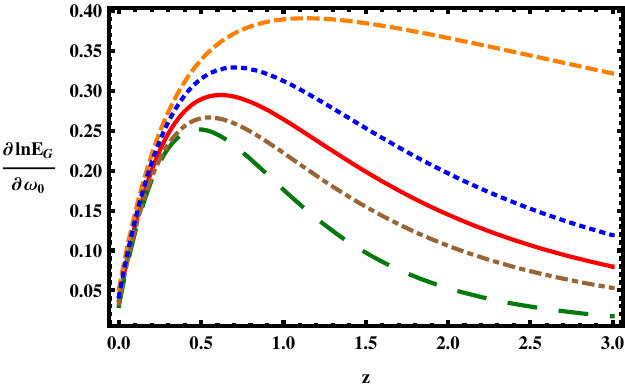} &
\includegraphics[width=0.3\linewidth]{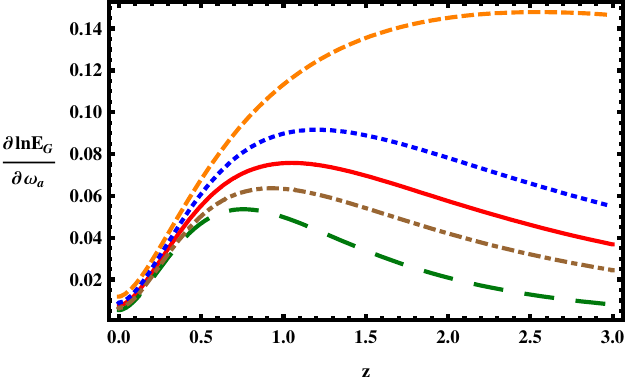} \\
\end{tabular}
\caption{ Logarithmic derivatives of the $\EG$ parameter with respect to (w.r.t) $\Omo$ (left), $\oo$ (middle), and $\oa$ (right) as a function of redshift for the different DE models. The representation of each curve corresponds to the same DE models used in Figure~\ref{fig2}.} \label{fig4}
\end{figure}

We emphasize that the $\EG$ statistic is considered here as an idealized theoretical observable. Our analysis is intended to assess the intrinsic sensitivity of $\EG$ to cosmological parameters, rather than to model a specific observational implementation involving galaxy bias, redshift-space distortions, lensing cross-correlations, scale cuts, or survey-dependent covariance matrices.  Incorporating survey-specific systematics and covariance modeling would
affect the achievable precision of $\EG$, but would not alter the
analytic degeneracy structure discussed here.
To assess the sensitivity of $\EG$ to cosmological parameters, we analyze its differential error
\begin{equation}
\fr{d\EG}{\EG} = \fr{\partial \ln \EG}{\partial \Omo} d \Omo + \fr{\partial \ln \EG}{\partial \oo} d \oo + \fr{\partial \ln \EG}{\partial \oa} d \oa \, . \label{dEG}
\end{equation}
The coefficients $\partial \ln \EG / \partial p$ (where $p \in \{\Omo, \oo, \oa\}$) for various DE models are presented as a function of redshift $z$ in Fig. \ref{fig4}. The curve representations for these coefficients are identical to those used in Figure \ref{fig2}. The overall redshift dependence of these sensitivity coefficients for $\EG$ exhibits qualitative similarities to those of the Hubble parameter $H(z)$. This resemblance can be understood from the approximate form of the growth factor in a flat universe, $\delta_g (z) \propto \fr{H(z)}{H_0} \int_{z}^{\infty} \fr{(1+z') dz'}{(H(z')/H_0)^3} \Omo(z')$ (where $\Omo(z) = \Omo (1+z)^3 / (H(z)/H_0)^2$), which shows a direct dependence on the Hubble parameter. Furthermore, the $\EG$ statistic involves derivatives of the Newtonian potentials, which are sourced by matter perturbations whose growth is governed by the expansion rate $H(z)$. Therefore, the sensitivity of $\EG$ to cosmological parameters inherits, to a significant extent, the sensitivity patterns observed in $H(z)$. While deviations from the simple growth factor formula exist for $\omega \neq -1.0$ or $-1/3$, these corrections are generally small, as discussed in the appendix. Consequently, $\EG$ provides complementary constraints on cosmological parameters, particularly those related to gravity and the growth of structure, when combined with geometric probes.

We scrutinize the sensitivity curves shown in figure \ref{fig4} as follows.
\begin{itemize}
    	\item \textbf{Sensitivity to $\Omo$ (Left Panel):} \\
The left panel of Figure~\ref{fig4} illustrates the sensitivity of the $\EG$ statistic to $\Omo$. $\frac{\partial \ln E_G}{\partial \Omo}$ is positive and generally increases with redshift for all considered DE models, indicating that $\EG$ becomes more sensitive to changes in the matter density at higher redshifts. Physically, $\EG$ is directly related to the ratio of gravitational potentials to peculiar velocity perturbations, both of which are strongly influenced by the abundance of matter. At higher redshifts, as the universe was more matter-dominated, variations in $\Omo$ exert a more pronounced effect on the growth of structure and, consequently, on $\EG$. The long-dashed (green) line ($\oo = -1.0, \oa = -0.8$) exhibits the highest sensitivity to $\Omo$ at high redshifts, suggesting that for DE models where DE density was less dominant in the past due to its rapid dilution (i.e., more negative $\oa$), the impact of matter density on $\EG$ measurements is most significant at early epochs. Conversely, the dashed (orange) line ($\oo = -1.0, \oa = +0.8$) displays the lowest sensitivity at high redshifts, implying that for models where DE becomes relatively more prominent at early times (positive $\oa$ means DE density was smaller in the past but its fractional contribution compared to matter decreases more slowly than for $\Lambda$CDM, leading to reduced matter dominance), the influence of matter density on $\EG$ is somewhat reduced at early times compared to other models. The $\Lambda$CDM (red solid), dotted (blue), and dot-dashed (brown) lines show intermediate sensitivities. This increasing sensitivity with redshift underscores that $\EG$ measurements at higher redshifts offer more stringent constraints on the matter density parameter.

A comparison between $\frac{\partial \ln H}{\partial \Omo}$ (Figure~\ref{fig2}, left panel) and $\frac{\partial \ln \EG}{\partial \Omo}$ (Figure~\ref{fig4}, left panel) reveals a similar increasing trend with redshift for all DE models. This signifies that both expansion-history probes ($H(z)$) and structure-growth probes ($\EG$) become more sensitive to $\Omo$ at higher redshifts, consistent with a more matter-dominated early universe. However, the magnitude of sensitivity differs significantly: $\frac{\partial \ln \EG}{\partial \Omo}$ is consistently larger than $\frac{\partial \ln H}{\partial \Omo}$ across the entire redshift range. This enhanced sensitivity of $\EG$ is attributed to its dependence on the growth factor $\delta(z)$, which amplifies the impact of $\Omo$ on observable quantities related to structure formation. While both $\ln H$ and $\ln \EG$ exhibit a roughly linear increase with $\ln(1+z)$ at high redshifts (as expected in a matter-dominated era), the pre-factor for $\ln \EG$ is larger, leading to a greater overall sensitivity to $\Omo$. Consequently, $\EG$ measurements, particularly at higher redshifts, can provide more stringent constraints on the present-day matter density compared to direct measurements of the Hubble parameter alone.

	\item \textbf{Sensitivity to $\oo$ (Middle Panel) and  $\oa$ (Right Panel):} \\
When comparing the sensitivities of the $\EG$ statistic and the Hubble parameter to the DE EOS parameters $\oo$ and $\oa$, we observe qualitatively similar redshift dependencies for both probes. For $\oo$, both $\frac{\partial \ln \EG}{\partial \oo}$ and $\frac{\partial \ln H}{\partial \oo}$ exhibit a non-monotonic behavior, initially increasing with redshift, peaking around $z \sim 0.5 - 1$, and then decreasing at higher redshifts. Similarly, for $\oa$, both $\frac{\partial \ln \EG}{\partial \oa}$ and $\frac{\partial \ln H}{\partial \oa}$ generally show an increasing trend at lower redshifts, followed by a peak or plateau at intermediate to high redshifts. This strong resemblance stems from the intrinsic link between $\EG$ and the growth rate of structure, which is fundamentally governed by the cosmic expansion history dictated by $H(z)$ and its evolution. Quantitatively, the magnitude of the sensitivities for $\EG$ tends to be comparable to, or slightly larger than, those for $H(z)$ across the redshift range. This suggests that $\EG$ measurements can provide complementary and potentially tighter constraints on the DE EOS parameters. The consistent relative ordering of the sensitivity curves for different DE models (characterized by varying $\oa$ values) between the two probes further indicates a similar qualitative response to the time evolution of DE in both the expansion rate and the growth of structure.
\end{itemize}

In summary, Figures \ref{fig2} and \ref{fig4} collectively demonstrate that the $\EG$ parameter exhibits a similar level of sensitivity to the time-varying DE component $\oa$ as the Hubble parameter $H(z)$. This implies that, within the framework of GR, even substantially improved $\EG$ measurements do not generically unlock new independent information directions for time-varying DE beyond those already accessible through $H(z)$. Nevertheless, the $\EG$ statistic remains a critical probe. Its importance lies in serving as an independent consistency check for GR on cosmological scales. Any significant deviation of the observed $\EG$ value from predictions based on standard gravity and the matter/DE content inferred from geometric probes would strongly indicate the necessity for alternative theories of gravity.

\section{Methodology and Analysis}
\label{sec:MAA}
\setcounter{equation}{0}

In this section, we describe the methodology used to analyze how various cosmological observables respond to variations in the underlying cosmological parameters. Our primary focus is on the sensitivity and degeneracy structure induced by different DE EOS parametrizations, rather than on data-driven parameter estimation. Throughout this section, our emphasis is on analytic sensitivity, local degeneracy geometry, and relative information content. Except where explicitly stated, no realistic observational likelihood, survey covariance, or forecasting analysis is performed.

We consider both geometric probes, such as the Hubble parameter $H(z)$ and the luminosity distance $D_L(z)$, and a structure-growth probe, the $\EG$ statistic. The analysis is carried out primarily within the CPL parametrization of the DE
EOS, $\omega(z)=\oo+\oa z/(1+z)$, and is later extended to an alternative phenomenological form, $\omega(z)=\oo+\oa\ln(1+z)$, in order to assess how the choice of parametrization redistributes parameter sensitivity.

\subsection{Fisher-Matrix Framework for Sensitivity Analysis}
\label{subsec:chi2}

To characterize how cosmological observables respond locally to variations in model parameters, we adopt a $\chi^2$-based formalism as a convenient theoretical framework. We emphasize that this formalism is introduced solely to define parameter sensitivities and degeneracy directions in parameter space. No actual $\chi^2$ minimization or parameter fitting to observational data is performed at this stage.

For an observable $f$ (such as $H(z)$, $D_L(z)$, or $\EG$) evaluated at a set of redshifts $z_i$ ($i=1,\dots,N$), one may formally write the $\chi^2$ function as
\begin{equation}
\chi^2 = \sum_{i=1}^{N} \frac{\bigl[f_i - f(z_i,\vec{p})\bigr]^2}{\sigma_i^2} \, 
\label{eq:chi2}
\end{equation}
where $f(z_i,\vec{p})$ denotes the theoretical prediction for a parameter vector $\vec{p}=(\Omo,\oo,\oa)$, and $\sigma_i$ represents the characteristic uncertainty associated with the observable at redshift $z_i$. In the present analysis, the quantities $f_i$ and $\sigma_i$ are introduced only to define a local metric in parameter space. They do not correspond to any specific observational dataset or survey.

Expanding the observable linearly around a fiducial parameter set $\vec{p}_{\rm fid}$, the curvature of the $\chi^2$ surface is characterized by the Fisher information matrix
\begin{equation}
\mathcal{F}_{pp'} = \frac{1}{2} \frac{\partial^2 \chi^2}{\partial p\,\partial p'} \Bigg|_{\vec{p}=\vec{p}_{\rm fid}} = \, \sum_{i=1}^{N}
g_i(z_i,\vec{p}_{\rm fid})\, g_i^{T}(z_i,\vec{p}_{\rm fid}) \, ,
\label{eq:fisher}
\end{equation}
with the sensitivity vectors
\begin{equation}
g_i(z_i,\vec{p}) = \frac{1}{\sigma_i} \frac{\partial f(z_i,\vec{p})}{\partial p} \Bigg|_{\vec{p}=\vec{p}_{\rm fid}} \, .
\end{equation}
Within this linear approximation, the Fisher matrix provides a compact description of how variations in different parameters project onto changes in the observable. Its inverse, $\mathcal{C}=\mathcal{F}^{-1}$, is interpreted here purely as a
diagnostic of local degeneracy structure, rather than as a quantitative covariance matrix for parameter inference.

\subsection{Alternative DE EOS Parametrizations: $\omega(z) = \oo + \oa \ln(1 + z)$}
\label{subsec:AEOSLnz}

While the CPL parametrization is used as our fiducial DE model, it is well known that different functional forms of $\omega(z)$ can redistribute parameter sensitivity in nontrivial ways. To explore this effect, we consider an alternative phenomenological parametrization, $\omega(z)=\oo+\oa\ln(1+z)$ (equivalently $\omega(a)=\oo-\oa\ln a$). For this parametrization, the Hubble parameter takes the form
\begin{equation}
H(z) = H_0 \sqrt{\Omega_m^0 (1+z)^3 + (1 - \Omega_m^0) (1+z)^{3+3\oo+\frac{3}{2} \oa \ln [1+z]} } \, .
\label{eq:Hz_alt}
\end{equation}
This parametrization is employed here as a low- to intermediate-redshift phenomenological description. It is not extrapolated to the recombination epoch, and no early-Universe or CMB distance priors are imposed. Our goal is to assess how alternative functional forms of $\omega(z)$ modify the distribution of sensitivity between $\oo$ and $\oa$ over the redshift range most relevant for late-time cosmological probes.

\begin{figure}
\centering
\vspace{1.5cm}
\begin{tabular}{ccc}
\includegraphics[width=0.3\linewidth]{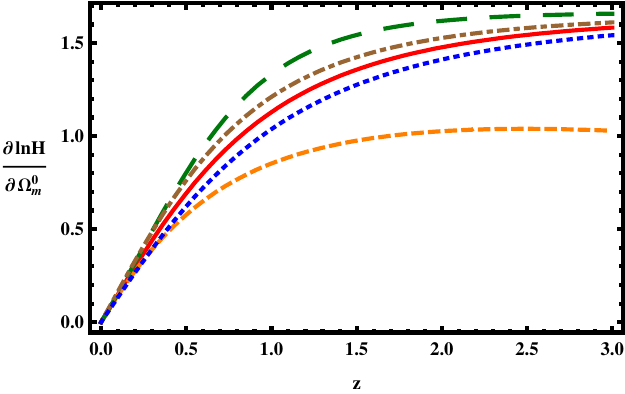} &
\includegraphics[width=0.3\linewidth]{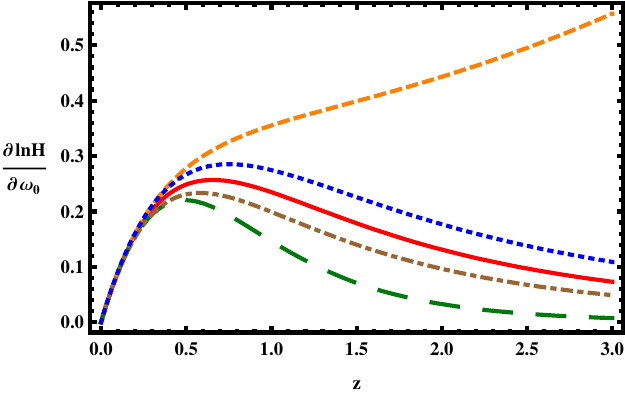} &
\includegraphics[width=0.3\linewidth]{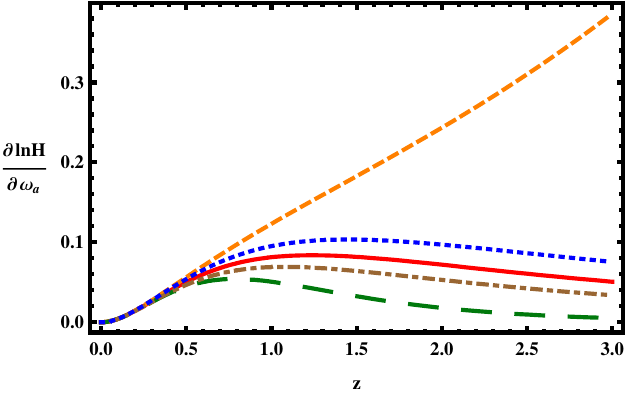} \\
\end{tabular}
\vspace{-0.5cm}
\caption{Logarithmic derivatives of the Hubble parameter w.r.t $\Omo$ (left), $\oo$ (middle), and $\oa$ (right) as a function of redshift for the different DE models when we use $\omega(z) = \oo + \oa \ln(1 + z)$. Note that the DE EOS used here is different from that in Figure \ref{fig2} (CPL parametrization), but the ($\oo$\,,\,$\oa$) values and the corresponding line styles employed for different DE models are identical to those used in Figure \ref{fig2}. }
\label{fig5}
\end{figure}

Figure~\ref{fig5} shows the logarithmic derivatives of $H(z)$ with respect to $\Omo$, $\oo$, and $\oa$ for this parametrization. Compared to the CPL case (Fig.~\ref{fig2}), the sensitivities to $\oo$ and $\oa$ exhibit more comparable magnitudes over a broad redshift range. This behavior reflects a redistribution of sensitivity geometry rather than a statement about physical superiority of one parametrization over another. Different parametrizations emphasize different combinations of parameters, even when applied to the same underlying expansion history.

As illustrated in the middle and right panels of Figure~\ref{fig5}, the sensitivities $\frac{\partial \ln H}{\partial \oo}$ and $\frac{\partial \ln H}{\partial \oa}$ for the $\omega(z) = \oo + \oa \ln(1 + z)$ parametrization display remarkably comparable magnitudes across a significant redshift range, especially at higher redshifts. This is a crucial contrast to the CPL parametrization (Figure~\ref{fig2}), where the sensitivity to $\oa$ is generally much weaker than that to $\oo$ (particularly for $\Lambda$CDM, where $\oa = 0$). This near parity in sensitivities to $\oo$ and $\oa$ in this alternative parametrization implies a more balanced contribution of both parameters to the overall uncertainty in $H(z)$. This characteristic potentially leads to improved and less degenerate constraints on the time-varying nature of the DE EOS. Specifically, the ability to probe $\oa$ with a sensitivity comparable to $\oo$ suggests that observational data can more effectively constrain the evolution of $\omega$ with redshift. This underscores the critical importance of choosing an appropriate parametrization of $\omega$ when investigating the properties of DE, as different parametrizations can yield significantly different sensitivities to the model parameters and, consequently, different constraining power from the same observational data. The more comparable sensitivities observed for $\oo$ and $\oa$ in the $\omega(z) = \oo + \oa \ln(1 + z)$ parametrization indicate its potential advantage over the CPL form for more precisely probing the dynamics of DE.  It is important to note, however, that while this parametrization offers improved sensitivity to $\oa$ at higher redshifts, the physical validity of any phenomenological DE parametrization, including this one, can be limited at very high redshifts. Such parametrizations are not typically derived from a fundamental theory of DE and may not accurately represent its behavior in the very early universe, where matter and radiation densities were dominant. Furthermore, mathematical behaviors such as potential divergences of the $\ln(1+z)$ term at extreme redshifts must be carefully considered. Observational constraints at very high redshifts are also generally limited, which means extrapolating these models significantly beyond the range of current data ($z \lesssim 2-3$) requires caution regarding their physical implications. We now provide a detailed analysis of the sensitivity of each observable to the cosmological parameters, starting with $H(z)$, followed by $D_L(z)$, and finally $\EG(z)$. This detailed scrutiny is crucial for understanding how different cosmic probes contribute to constraining the properties of DE.

\subsection{Fiducial Cosmological Model and Analytic Solutions}
\label{subsec:FCMAS}

For the special case of a constant DE EOS $\omega$, exact analytic solutions exist for several key cosmological quantities in a spatially flat universe. The Hubble parameter is given by
\begin{equation}
H(z) = H_0 \sqrt{ \Omo (1+z)^3 + (1-\Omo)(1+z)^{3(1+\omega)}} \, .
\label{Hcon}
\end{equation}
Closed-form expressions for the conformal time, comoving distance, and linear growth factor can be written in terms of hypergeometric functions~\cite{Lee:2009gb,Lee:2009ft,Lee:2009if}.

These analytic solutions are used to validate and interpret the numerical sensitivity calculations presented in this work. They are not employed as inputs to parameter fitting or likelihood analyses. The conformal time, related to the Hubble parameter by $d \eta = -\fr{dz}{H(z)}$, can be expressed analytically using hypergeometric functions
\begin{align}
\eta(z) &= \int_{z}^{\infty} \fr{dz'}{H(z')} = \fr{2 (1+z)}{H(z)} \, \fr{1}{\sqrt{\Omz}} \, F \Bigl[\fr{1}{2}\, , -\fr{1}{6\omega}\, , 1 - \fr{1}{6\omega}\, , -\fr{\rhode (z)}{\rhom (z)} \Bigr] \nonumber \\ &= \fr{2 (1+z)}{H(z)} \, F \Bigl[\fr{1}{2}\, , 1\, , 1 - \fr{1}{6\omega}\, , 1 - \Omz \Bigr] \, . \label{eta}
\end{align}
where $\Omega(z) = \fr{\Omo (1+z)^3}{H(z)^2/H_0^2}$ is the redshift-dependent matter density parameter. The comoving distance $r(z)$, defined as $r(z) = \int_{0}^{z} \fr{dz'}{H(z')}$, is then simply the difference in conformal time between $z=0$ and $z$ ($r(z) = \eta(0) - \eta(z)$)
\begin{equation}
r(z) = \fr{2}{H_0} \, F \Bigl[\fr{1}{2}\, , 1\, , 1 - \fr{1}{6\omega}\, , 1 - \Omo \Bigr] -  \fr{2 (1+z)}{H(z)} \, F \Bigl[\fr{1}{2}\, , 1\, , 1 - \fr{1}{6\omega}\, , 1 - \Omz \Bigr] \, .\label{r}
\end{equation}
The sub-horizon-scale linear perturbation equation governing the growth of density perturbations $\delta$ with respect to the scale factor $a$ (or redshift $z$) is given by
\begin{equation}
\fr{d^2 \delta}{dz^2} + \Biggl( \fr{d \ln H}{d z} - \fr{1}{1+z} \Biggr) \fr{d \delta}{d z} - \fr{4 \pi G \rho_{m}}{((1+z) H)^2} \delta = 0 \, .
\label{dzdelta}
\end{equation}
The exact analytic growing mode solution $\delta_g(z)$ for this equation with a constant $\omega$ is well-established \cite{Lee:2009gb,Lee:2009ft,Lee:2009if} and involves hypergeometric functions
\begin{align}
\delta (z) &= c_{1} \Omega^{\fr{3 \omega -1}{6 \omega}} (1+z)^{\fr{1 - 3 \omega}{2}} F [\fr{\omega -1}{2} , \fr{3 \omega + 2}{6 \omega} , \fr{9 \omega - 1}{6 \omega} , - \Omega (1+z)^{-3 \omega}] \nonumber \\ & \, + \, c_{2} F[-\fr{1}{3\omega}, \fr{1}{2 \omega}, \fr{3 \omega + 1}{6} , -\Omega (1+z)^{-3 \omega}] \, , \label{deltask}
\end{align}
where $\Omega = \fr{\Omo}{1-\Omo}$ and $F$ is the hypergeometric function. The ratio of the coefficients $c_1$ and $c_2$ is determined by initial conditions, often evaluated at a high redshift $z_i$
\begin{align}
&& \fr{c_1}{c_2} \Bigl(z_{i}, \Omega, \omega \Bigr) = 2 (1 + z_{i})^{\fr{3\omega - 1}{2}} \Omega^{\fr{1 - 3 \omega}{6\omega}} (9\omega -1) \Biggl( -(1+3\omega) \times \nonumber \\ && F \Bigl[-\fr{1}{3\omega},\fr{1}{2\omega}, \fr{3 \omega + 1}{6\omega},-Y_i \Bigr] \nonumber + 3 Y_i F \Bigl[\fr{3\omega - 1}{3\omega}, \fr{2 \omega + 1}{2\omega}, \fr{9 \omega + 1}{6\omega}, -Y_i\Bigr] \Biggr) \nonumber \\ && \Bigg/ 3(3\omega+1)(\omega-1) \Biggl( Y_i (3\omega+2) F \Bigl[\fr{3 \omega - 1}{2\omega} , \fr{9 \omega + 2}{6\omega} , \fr{15 \omega - 1}{6\omega} , -Y_i \Bigr] \nonumber \\ && + (1-9\omega) F \Bigl[-\fr{1 + \omega}{2\omega} , \fr{3 \omega + 2}{6\omega}, \fr{9 \omega - 1}{6\omega}, -Y_i \Bigr] \Biggr) \, , \label{c1c2} 
\end{align}
where $Y_i = (1+z_{i})^{-3\omega}\Omega$.

\subsection{Illustrative Parameter-Recovery Exercise}
\label{subsec:PCA}

To illustrate how the analytic sensitivity and degeneracy structures discussed in the previous sections manifest in a conventional parameter-space representation, we present an illustrative parameter-recovery exercise based on mock $H(z)$ data. This exercise is pedagogical in nature and is not intended as a realistic forecast or as a source of observational constraints.

We construct a formal $\chi^2_H$ using mock Hubble-parameter measurements with assumed Gaussian errors,
\begin{equation}
\chi^2_H = \sum_{i=1}^{N} \frac{\bigl[H_{\rm obs}(z_i)-H_{\rm th}(H_0,\Omo,p_j;z_i)\bigr]^2}{\sigma_H^2(z_i)} \, , \label{eq:chi2_H}
\end{equation}
where $p_j=(\oo,\oa)$. The mock data $H_{\rm obs}(z_i)$ and uncertainties $\sigma_H(z_i)$ are chosen to be representative of current $H(z)$ measurements but do not correspond to any specific survey.  The purpose of this construction is solely to visualize relative degeneracy directions and sensitivity hierarchies implied by the analytic derivatives discussed in Sec.~\ref{sec:COTS}. No claim is made regarding quantitative parameter constraints.

To illustrate the role of nuisance parameters, we optionally marginalize over the Hubble constant $H_0$ assuming a Gaussian prior,
\begin{equation}
\pi_H(H_0) \propto \exp\!\left[-\frac{(H_0-H_0^{\rm obs})^2}{2\sigma_{H_0}^2} \right] ,
\end{equation}
leading to the marginalized statistic
\begin{equation}
\tilde{\chi}^2(\Omo,p_j) = -2\ln \!\left[ \int dH_0\, \pi_H(H_0)\, \exp\!\left(-\frac{\chi_H^2}{2}\right) \right] .
\label{eq:marginalized_chi2}
\end{equation}
For a linear dependence of $H_{\rm th}$ on $H_0$, this marginalization can be performed analytically.

\begin{figure*}
\centering
\begin{tabular}{ccc}
\includegraphics[width=0.3\linewidth]{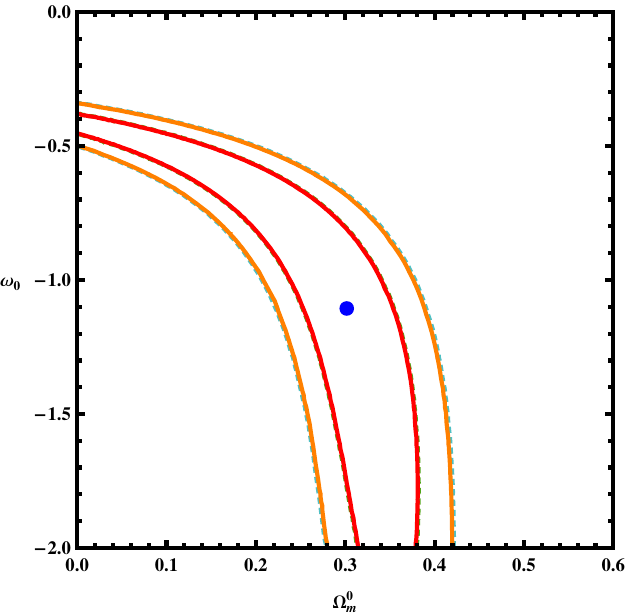} &
\includegraphics[width=0.3\linewidth]{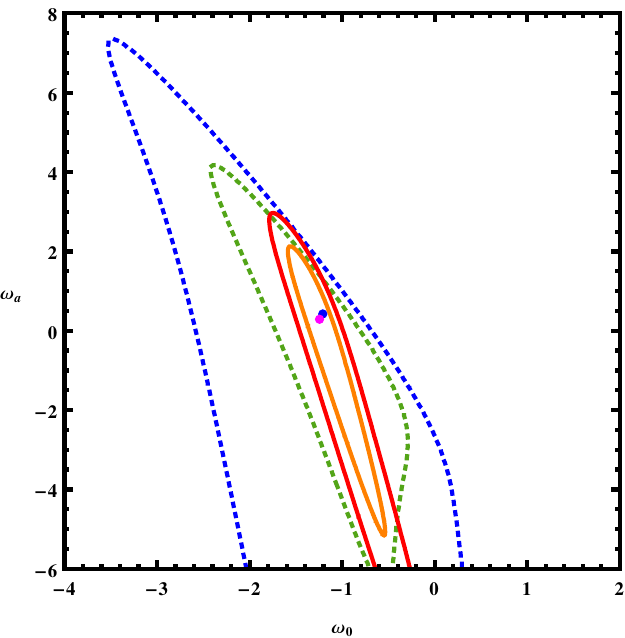} &
\includegraphics[width=0.3\linewidth]{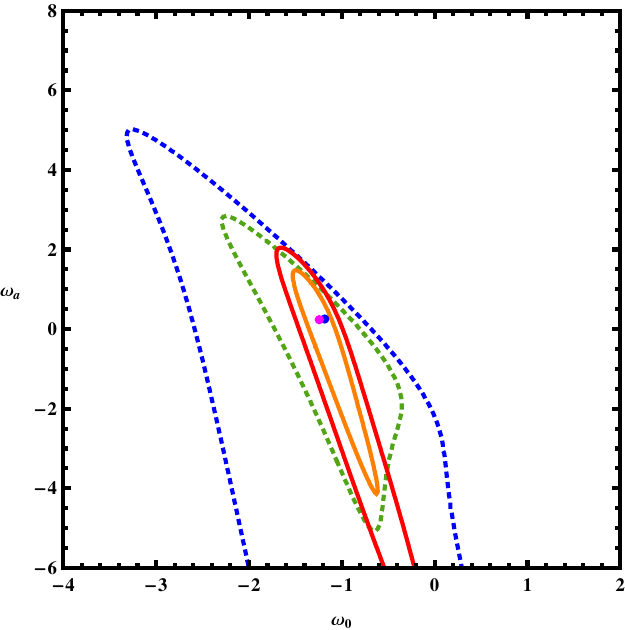} \\
\end{tabular}
\caption{
Illustrative parameter-space contours derived from the mock $H(z)$ exercise.
\textbf{Left}: Degeneracy structure in the $(\Omo,\oo)$ plane with and without
marginalization over $H_0$.
\textbf{Middle}: Degeneracy directions in the $(\oo,\oa)$ plane for the CPL
parametrization, shown for $H(z)$ alone and for a combined illustrative
$H+H_{\rm BAO}$ sensitivity.
\textbf{Right}: Corresponding degeneracy structure for the alternative
parametrization $\omega(z)=\oo+\oa\ln(1+z)$.
In all panels, the fiducial $\Lambda$CDM point is marked for reference.
}
\label{fig6}
\end{figure*}

Figure~\ref{fig6} visualizes how the analytic sensitivity structure translates into parameter-space geometry. The left panel shows that marginalization over $H_0$ primarily stretches the allowed region along a direction already weakly constrained by $H(z)$, consistent with the hierarchy of sensitivities identified in Fig.~\ref{fig2}. The middle and right panels compare the CPL and logarithmic parametrizations of the DE EOS. For the CPL case, the contours in the $(\oo,\oa)$ plane remain highly elongated, reflecting the suppressed sensitivity to $\oa$ discussed earlier. In contrast, the $\omega(z)=\oo+\oa\ln(1+z)$ parametrization yields more compact and symmetric contours, indicating a more balanced distribution of sensitivity between $\oo$ and $\oa$. This improvement should be interpreted as a redistribution of sensitivity geometry rather than as evidence for stronger physical constraints. Different parametrizations project the same underlying information content onto different parameter combinations.

Overall, this illustrative exercise reinforces the central message of this work: the apparent constraining power on time-varying DE parameters depends sensitively on the chosen parametrization and on how observational information is geometrically mapped in parameter space.

\subsection{Comparison of Sensitivity Between $H(z)$ and $\EG$}
\label{subsec:CSHEG}

Finally, we compare the sensitivity patterns of the expansion-history probe $H(z)$ and the growth-related $\EG$ statistic. Although $\EG$ is constructed from structure-growth observables, its dependence on the linear growth rate and the Newtonian potentials ties it closely to the background expansion history governed by $H(z)$.

In linear perturbation theory within GR, the $\EG$ statistic can be written schematically as $\EG \propto \Omo/f(z)$, where $f(z)=d\ln\delta/d\ln a$ is the linear growth rate. Since the growth factor $\delta(z)$ is itself controlled by the expansion rate, the response of $\EG$ to variations in cosmological parameters inherits much of the same redshift dependence as that of $H(z)$. For slowly varying DE models, approximate analytic solutions of the growth equation show that the dominant parameter dependence of $f(z)$ is set by $H(z)$ and its first derivative. As a result, the logarithmic derivatives $\partial\ln\EG/\partial p$ and $\partial\ln H/\partial p$ (with $p=\{\Omo,\oo,\oa\}$) exhibit similar hierarchies and redshift trends, as demonstrated in Figs.~\ref{fig2} and \ref{fig4}. This similarity implies that, within GR, improved measurements of $\EG$ primarily reinforce information already encoded in the expansion history, rather than introducing a new, independent sensitivity direction for time-varying DE. Consequently, the main cosmological role of $\EG$ is not to sharpen constraints on $(\oo,\oa)$ beyond those achievable with geometric probes, but to serve as an internal consistency test of gravity on cosmological scales.

Any statistically significant deviation between the $\EG$-based sensitivity structure and that inferred from $H(z)$ would therefore provide a strong and model-independent indication of modified gravity or non-standard clustering properties of DE.

\section{Interpretation of Sensitivity Structure and Reliability of Dark-Energy Inference}
\label{sec:sensitivities}

In this section, we clarify how the analytic sensitivity structures derived in Sections~\ref{sec:COTS} and \ref{sec:MAA} inform the interpretation of cosmological parameter constraints, particularly in the context of time-varying DE. Our emphasis is not on deriving new observational limits, but on understanding why existing analyses frequently exhibit weak, one-sided, or prior-dependent constraints on the evolution parameter $\oa$.

Throughout this section, no new observational likelihoods are constructed and no cosmological data are reanalyzed. Instead, we focus on the internal geometry of parameter space and on how parameter degeneracies respond to external priors. This approach is intended to complement, rather than compete with, data-driven analyses.

\subsection{From Sensitivity Hierarchies to Apparent Constraints}
\label{subsec:sensitivity_to_constraints}

The sensitivity analysis presented in Section~\ref{sec:COTS} revealed a clear hierarchy in how geometric observables respond to cosmological parameters. For both $H(z)$ and $D_L(z)$, the logarithmic derivatives with respect to $\Omo$ dominate over those with respect to $\oo$, while the sensitivity to $\oa$ is generically the weakest, often by an order of magnitude.

This hierarchy has immediate consequences for parameter inference. When an observable exhibits intrinsically weak sensitivity to a given parameter, even modest measurement uncertainties or external priors can dominate the inferred value of that parameter. As a result, apparent constraints on $\oa$ may largely reflect the projection of stronger constraints on $(\Omo, \oo)$ onto a nearly degenerate direction, rather than a genuine sensitivity to DE evolution itself.

These considerations motivate a careful distinction between the numerical tightness of reported constraints and their structural robustness. In the remainder of this section, we illustrate this distinction explicitly.

\subsection{Illustrative Response of $(\oo,\oa)$ to an External $\Omo$ Prior}
\label{subsec:degeneracy_response}

To visualize how the sensitivity structure manifests itself in a familiar parameter-space representation, we perform an illustrative Monte Carlo experiment based on mock luminosity-distance data. This exercise is pedagogical in nature and is not intended to model any specific survey, nor to provide forecasts or observational constraints.

The sole purpose of this experiment is to demonstrate how an external prior on $\Omo$ propagates into the $(\oo\,,\oa)$ subspace in the presence of strong degeneracies inherent to the CPL parametrization.

We adopt a fiducial flat $\Lambda$CDM cosmology with $(\Omo,\oo,\oa) = (0.3, -1, 0)$ and generate mock distance-modulus data from the corresponding luminosity distance. The mock sample consists of a smoothly distributed set of supernovae over the redshift range $0.01 \le z \le 3$, with a redshift-dependent scatter chosen to be representative of current distance-modulus uncertainties. The precise number of supernovae and the detailed form of the redshift distribution are not critical for the conclusions below, which depend only on the relative sensitivity structure of $D_L(z)$.

For each mock realization, we minimize a total $\chi^2$ of the form
\begin{equation}
\chi^2_{\rm tot} = \chi^2_{\rm SN} + \left(\frac{\Omo-\Omo^{\rm prior}}{\sigma_{\Omo}} \right)^2 ,
\end{equation}
where the second term represents a Gaussian prior on $\Omo$. The prior width is held fixed, while the prior mean $\Omo^{\rm prior}$ is varied systematically. This procedure isolates the response of $(\oo, \oa)$ to shifts in the assumed matter density.

\subsection{Bias Directions in the $(w_0, w_a)$ Plane}
\label{subsec:bias_directions}

\begin{figure}[t]
\centering
\includegraphics[width=0.9\linewidth]{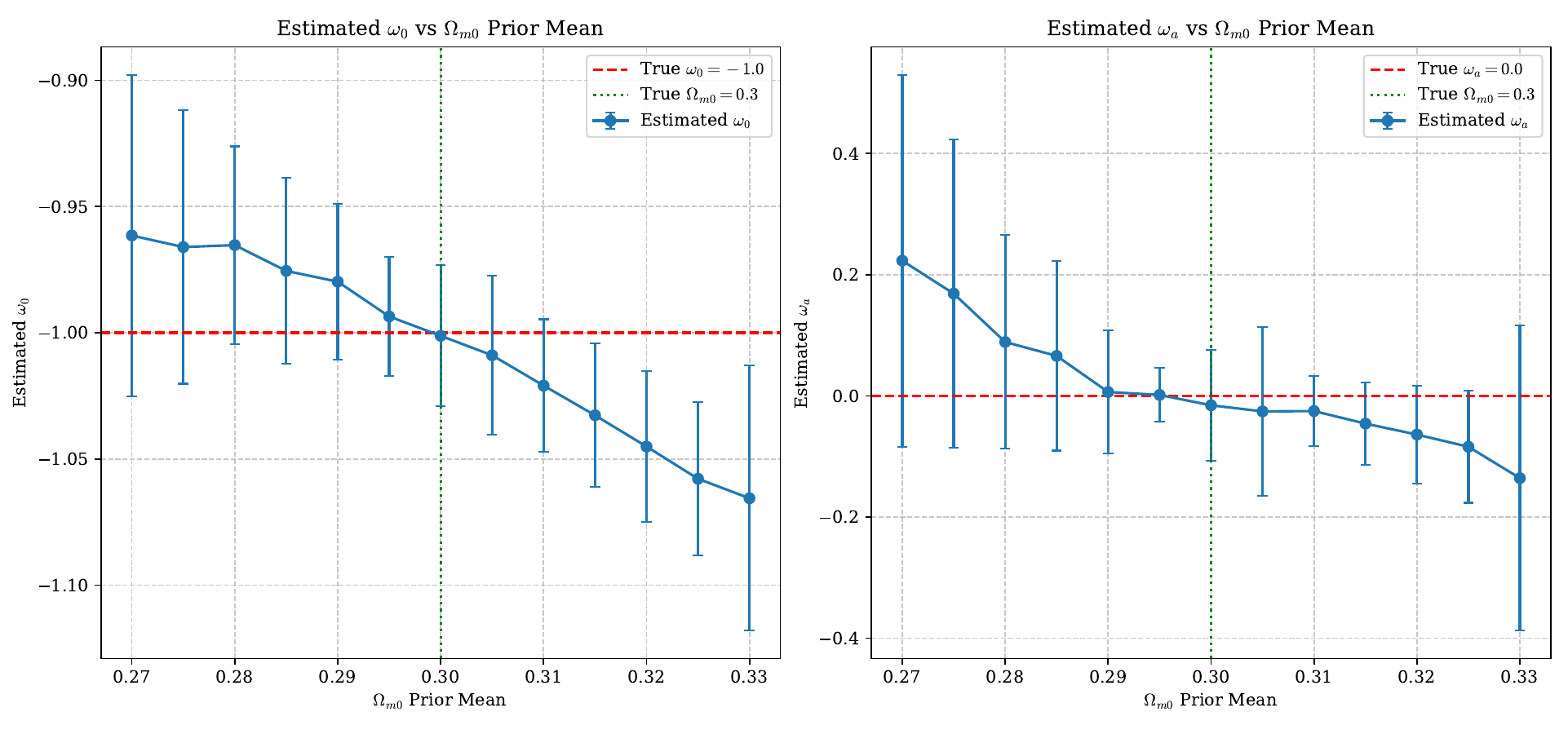}
\caption{Recovered mean values of $\oo$ (left) and $\oa$ (right) as a function of the assumed prior mean $\Omo^{\rm prior}$.
Points show the ensemble-averaged best-fit values obtained from $N_{\rm mock}=100$ realizations, while vertical error bars indicate the $1\sigma$ scatter across the mocks. The horizontal dashed lines mark the fiducial input values $(\oo, \oa)=(-1,0)$, and the vertical dotted line indicates the true $\Omo=0.30$.}
\label{fig7}
\end{figure}

Figure~\ref{fig7} displays the ensemble-averaged best-fit values of $\oo$ and $\oa$ as functions of the imposed  $\Omo$ prior mean. The true fiducial values are indicated for reference.

Several generic features are immediately apparent. First, both $\oo$ and $\oa$ exhibit systematic shifts as the prior mean on
$\Omo$ is varied. Second, these shifts follow well-defined directions in parameter space, reflecting the underlying degeneracy between matter density and DE parameters. Physically, increasing $\Omo$ enhances the decelerating influence of matter on the expansion history. To compensate and reproduce the same luminosity-distance data, the DE sector must provide stronger late-time acceleration, driving the best-fit values of $\oo$ and $\oa$ toward more negative values. This behavior is a direct manifestation of the degeneracy directions encoded in the sensitivity kernels derived earlier. Importantly, these trends arise even though the underlying mock data are generated from $\Lambda$CDM. They therefore illustrate how biased estimates of $(\oo, \oa)$ can emerge purely from prior mis-centering, without any true time variation of DE.

\subsection{Controlled Monte Carlo Illustration of Degeneracy and Error Propagation}
\label{subsec:mock_sne}

The purpose of this subsection is not to derive new observational constraints, but to provide a controlled and transparent numerical illustration of how degeneracies between $\Omo$ and the DE EOS parameters $(\oo, \oa)$ translate into bias and uncertainty propagation. All results shown below are obtained from synthetic supernova data generated from a fiducial cosmology and should be interpreted purely as a pedagogical exercise.

\subsubsection{Numerical setup and mock data generation}

To make the construction of Figs.~\ref{fig7}--\ref{fig8} fully reproducible, we summarize the exact numerical choices adopted in the Monte Carlo simulation. No real observational data or survey likelihoods are used at any stage. The entire analysis is based on mock Type-Ia supernova samples generated from a fiducial flat $\Lambda$CDM cosmology.

\begin{table}[t]
\centering
\caption{Numerical configuration used to generate Figs.~\ref{fig7} and~\ref{fig8}. All quantities are fixed deterministically, except for the Gaussian noise realizations applied to the mock distance moduli.}
\label{tab:mock_sne_numeric}
\begin{tabular}{lc}
\hline\hline
Item & Choice used in this work \\
\hline
Fiducial cosmology & $(\Omo, \oo, \oa)=(0.30,-1,0,70)$ \\ Redshift range & $z\in[0.01,3.0]$ \\ Number of supernovae & $N_{\rm SN}=1500$ \\ Magnitude error model & $\sigma_\mu(z)=0.15+0.05z$ \\ Number of mock realizations & $N_{\rm mock}=100$ \\ $\Omo$ prior mean & $0.27$--$0.33$ in steps of $0.005$ \\ $\Omo$ prior width & $\sigma_{\Omo}=0.01$ \\ Fit parameters & $(\Omo, \oo, \oa)$ \\ Optimization method & bounded $\chi^2$ minimization \\
\hline\hline
\end{tabular}
\end{table}

\subsubsection{Bias induced by the $\Omo$ prior}

Figure~\ref{fig7} shows how a mis-centered prior on $\Omo$ induces systematic shifts in the recovered DE parameters. As the assumed prior mean deviates from the true value, the best-fit $\oo$ and $\oa$ move coherently along the intrinsic degeneracy directions. This behavior reflects the limited ability of luminosity-distance data to independently disentangle matter density from DE evolution.

\subsubsection{Propagation of prior uncertainty into $\oo$ and $\oa$}

\begin{figure}[t]
\centering
\includegraphics[width=0.7\linewidth]{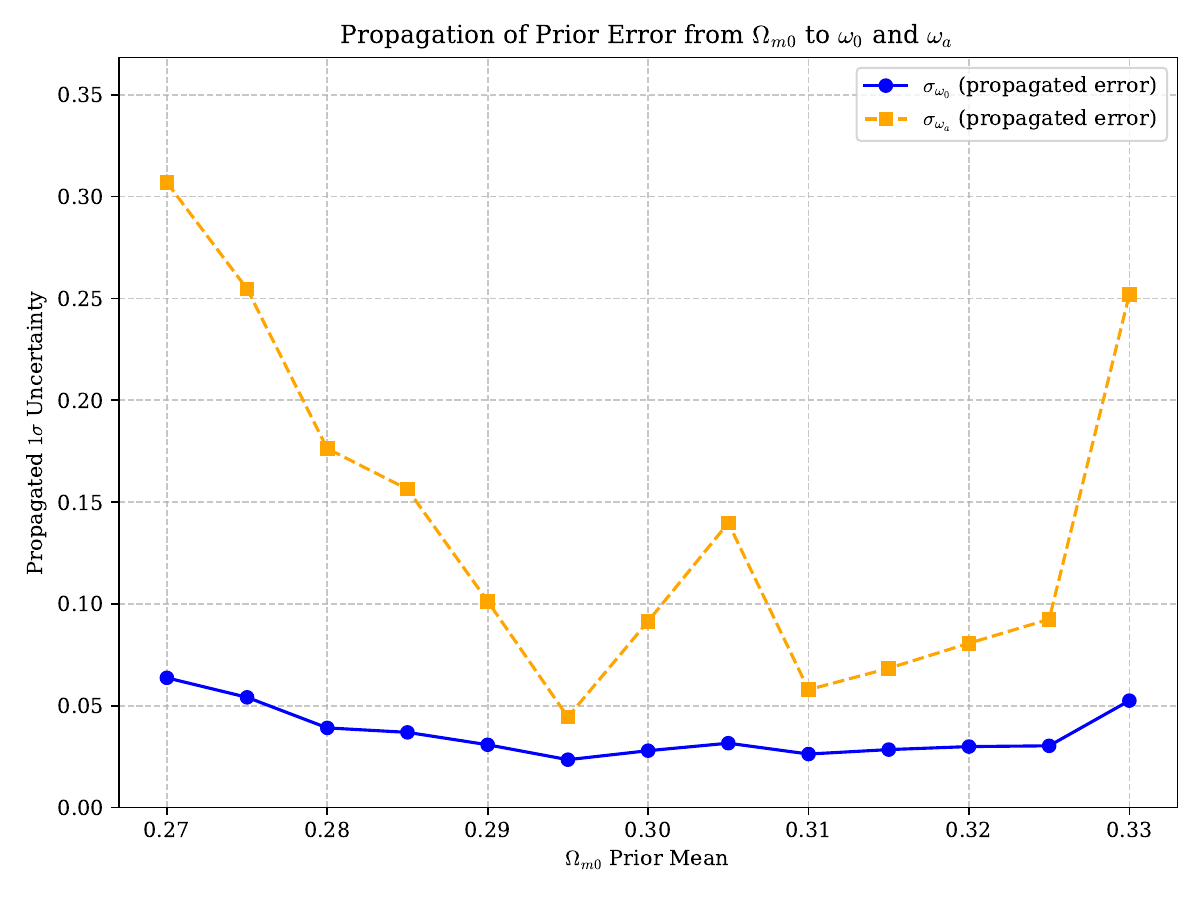}
\caption{Propagated $1\sigma$ uncertainties of $\oo$ (solid blue) and $\oa$ (dashed orange) as a function of the assumed prior mean $\Omo^{\rm prior}$. Each curve represents the standard deviation of the recovered parameters across the mock ensemble at fixed prior mean.}
\label{fig8}
\end{figure}

Figure~\ref{fig8} quantifies how uncertainty in $\Omo$ propagates into the DE sector.

Across the entire prior range, the uncertainty $\sigma(\oa)$ remains substantially larger than $\sigma(\oo)$, reflecting the intrinsically weaker sensitivity of geometric observables to time variation in the EOS. Even when the prior is correctly centered, the residual uncertainty on $\oa$ remains large, underscoring the difficulty of constraining $\oa$ robustly with distance data alone.

Taken together, Figs.~\ref{fig7} and~\ref{fig8} provide a controlled numerical demonstration of how parameter degeneracies and prior assumptions can both bias and broaden constraints on $(\oo, \oa)$, independently of any survey-specific systematics or likelihood choices.

\subsection{Implications for Interpreting Time-Varying Dark Energy}
\label{subsec:implications}

The illustrative experiments presented above reinforce a central message of this work. In the presence of strong degeneracies and weak intrinsic sensitivity, apparent constraints on $\oa$ should be interpreted with caution. Numerical tightness alone does not guarantee robustness, particularly when constraints depend sensitively on external priors or parametrization choices.

Rather than favoring or excluding time-varying DE, this analysis highlights the importance of distinguishing genuine sensitivity to DE dynamics from effects induced by parameter geometry. In this sense, sensitivity diagnostics provide a framework for interpreting data-driven results reliably, rather than replacing them.

\section{Conclusions}
\label{sec:Con}
\setcounter{equation}{0}

In this work, we have systematically examined how key cosmological observables encode information about dark-energy parameters, with particular emphasis on the geometric and growth-based probes commonly used in late-time cosmology. Rather than presenting new observational constraints, our analysis has focused on understanding the intrinsic sensitivity structure of these observables and the implications this structure has for the robustness of dark-energy inference.

A central result of our study is the clear hierarchy in parameter sensitivities. Across all observables considered---including $H(z)$, $D_L(z)$, and the $\EG$ statistic---the response to the matter density $\Omo$ (and to a lesser extent $H_0$) dominates over the response to the dark-energy equation-of-state parameters $(\oo, \oa)$. In particular, the sensitivity to the time-variation parameter $\oa$ is consistently the weakest, often by an order of magnitude. This hierarchy implies that precise knowledge of $\Omo$ is a necessary precondition for any robust attempt to constrain time-varying dark energy. Physically, this reflects the dominant role of matter--dark-energy competition in shaping the recent expansion history and the transition to cosmic acceleration.

Our analysis further clarifies the relationship between expansion-history probes and growth-based observables. Within General Relativity, we find that the $\EG$ statistic exhibits sensitivity patterns closely aligned with those of $H(z)$, indicating that improved growth measurements primarily reinforce information already encoded in the background expansion. The distinctive value of $\EG$ therefore lies less in opening new sensitivity directions for dark energy and more in providing a powerful consistency test of gravity on cosmological scales.

A key implication of these sensitivity patterns is that apparently tight or one-sided constraints on $w_a$ do not necessarily imply genuine sensitivity to dark-energy evolution. Through controlled Monte Carlo experiments with mock supernova data, we have demonstrated explicitly how external priors---particularly on $\Omo$---can induce systematic shifts and large uncertainty propagation in the $(\oo, \oa)$ subspace, even when the underlying cosmology is exactly $\Lambda$CDM. These effects arise purely from parameter degeneracies and are independent of survey-specific systematics or likelihood choices. They therefore highlight the need to distinguish numerical tightness from structural robustness when interpreting claims of time-varying dark energy.

Importantly, the qualitative conclusions of this work are rooted in the geometry of parameter space and the analytic sensitivity structure of the observables. As such, they are robust to moderate extensions of the cosmological parameter set, including spatial curvature ($\Omega_k \neq 0$) or a free summed neutrino mass ($\sum m_\nu$). While such extensions would quantitatively weaken constraints by enlarging the degenerate subspace, they do not introduce new independent sensitivity directions capable of fundamentally altering our conclusions.

Our results also underscore the critical role of parametrization. Within the widely used CPL form, the intrinsically weak sensitivity to $\oa$ severely limits the detectability of dark-energy evolution. We have shown that alternative phenomenological parametrizations, such as $\omega(z)=\oo+\oa\ln(1+z)$, redistribute sensitivity more evenly between parameters and can substantially enhance the constraining power on time variation. This highlights that progress in dark-energy studies depends not only on improved data quality but also on informed theoretical choices in how dark energy is parametrized.

In summary, the primary contribution of this work is to provide a coherent interpretive framework for understanding why current and near-future analyses often struggle to deliver robust constraints on time-varying dark energy. By linking observed parameter constraints to the underlying sensitivity and degeneracy structure of cosmological observables, we offer a diagnostic tool for assessing the reliability of dark-energy inferences. Future advances will require a concerted strategy that combines high-precision measurements of fundamental parameters such as $\Omo$ and $H_0$, diverse and complementary observational probes, and carefully chosen parametrizations designed to maximize sensitivity to physically meaningful degrees of freedom. Only through such an approach can the true nature of cosmic acceleration be reliably unraveled.

\section*{Acknowledgments}
SL is supported by the National Research Foundation of Korea (NRF), funded by the Ministry of Education (Grant No. NRF-RS202300243411).  

\appendix
\section{Analytic expressions for $H(z)$ and derivatives of $\ln H$}
\label{app:dlnH}

In this appendix we provide the verified analytic expressions for $H(z)$ and the exact derivatives $\partial \ln H/\partial \Omo$, $\partial \ln H/\partial \oo$, and $\partial \ln H/\partial \oa$ used to generate Figs.~\ref{fig2}--\ref{fig5}. Throughout we assume a spatially flat universe and neglect radiation in the redshift range of interest, so that $\Ode=1-\Omo$ and
\begin{equation}
E^2(z)\equiv \frac{H^2(z)}{H_0^2} = \Omo (1+z)^3 + (1-\Omo)\,f_{\rm DE}(z),
\label{eq:Ez2_general}
\end{equation}
where $f_{\rm DE}(z)=\rho_{\rm DE}(z)/\rho_{\rm DE,0}$ encodes the dark-energy evolution. For any parameter $p \in \{\Omo,\oo,\oa\}$, we use
\begin{equation}
\frac{\partial \ln H}{\partial p} = \frac{1}{2}\,\frac{1}{E^2(z)}\,\frac{\partial E^2(z)}{\partial p}.
\label{eq:dlnH_general}
\end{equation}

\subsection{CPL parameterization}
\label{app:dlnH_cpl}

For the CPL form $\omega(z)=\oo+\oa\,\dfrac{z}{1+z}$, the dark-energy evolution is
\begin{equation}
f_{\rm DE}^{\rm CPL}(z) = (1+z)^{3(1+\oo+\oa)}\exp\!\left[-3\oa\frac{z}{1+z}\right]. \label{eq:fde_cpl}
\end{equation}
Using Eq.~\eqref{eq:Ez2_general}, we have
\begin{equation}
E^2_{\rm CPL}(z) = \Omo(1+z)^3 + (1-\Omo)\,f_{\rm DE}^{\rm CPL}(z).
\label{eq:Ez2_cpl}
\end{equation} The required derivatives of $\ln H$ are then
\paragraph{Derivative with respect to $\Omo$.}
\begin{equation}
\frac{\partial E^2_{\rm CPL}}{\partial \Omo} = (1+z)^3 - f_{\rm DE}^{\rm CPL}(z),
\qquad
\boxed{ \frac{\partial \ln H}{\partial \Omo} = \frac{(1+z)^3 - f_{\rm DE}^{\rm CPL}(z)}{2\,E^2_{\rm CPL}(z)} }.
\label{eq:dlnH_dOm_cpl}
\end{equation}

\paragraph{Derivatives with respect to $\oo$ and $\oa$.}
It is convenient to note that
\begin{equation}
\frac{\partial f_{\rm DE}^{\rm CPL}}{\partial \oo} = 3\ln(1+z)\, f_{\rm DE}^{\rm CPL}(z),
\qquad
\frac{\partial f_{\rm DE}^{\rm CPL}}{\partial \oa} = 3\!\left[\ln(1+z)-\frac{z}{1+z}\right] f_{\rm DE}^{\rm CPL}(z).
\label{eq:dfde_cpl}
\end{equation}
Therefore
\begin{equation}
\frac{\partial E^2_{\rm CPL}}{\partial \oo} = (1-\Omo)\,3\ln(1+z)\, f_{\rm DE}^{\rm CPL}(z),
\label{eq:dE2_dw0_cpl}
\end{equation}
\begin{equation}
\frac{\partial E^2_{\rm CPL}}{\partial \oa} = (1-\Omo)\,3\!\left[\ln(1+z)-\frac{z}{1+z}\right] f_{\rm DE}^{\rm CPL}(z).
\label{eq:dE2_dwa_cpl}
\end{equation}
Combining with Eq.~\eqref{eq:dlnH_general} yields
\begin{equation}
\boxed{\frac{\partial \ln H}{\partial \oo} = \frac{3(1-\Omo)\,\ln(1+z)\, f_{\rm DE}^{\rm CPL}(z)}{2\,E^2_{\rm CPL}(z)}},
\qquad
\boxed{\frac{\partial \ln H}{\partial \oa} = \frac{3(1-\Omo)\left[\ln(1+z)-\frac{z}{1+z}\right] f_{\rm DE}^{\rm CPL}(z)}{2\,E^2_{\rm CPL}(z)}}.
\label{eq:dlnH_dw0dwa_cpl}
\end{equation}

\subsection{$\ln(1+z)$ parameterization}
\label{app:dlnH_log}

For the logarithmic form $\omega(z)=\oo+\oa\,\ln(1+z)$, the evolution is obtained from
\begin{equation}
f_{\rm DE}(z) = \exp\!\left[3\int_{0}^{z}\frac{1+\omega(z')}{1+z'}\,dz'\right].
\label{eq:fde_def}
\end{equation}
Since $\int_0^z \dfrac{dz'}{1+z'}=\ln(1+z)$ and $\int_0^z \dfrac{\ln(1+z')}{1+z'}dz'=\dfrac{1}{2}\ln^2(1+z)$, we obtain
\begin{equation}
f_{\rm DE}^{\ln}(z) = (1+z)^{3(1+\oo)}\exp\!\left[\frac{3}{2} \oa \ln^2(1+z)\right].
\label{eq:fde_log}
\end{equation}
Thus
\begin{equation}
E^2_{\ln}(z) = \Omo(1+z)^3 + (1-\Omo)\,f_{\rm DE}^{\ln}(z). \label{eq:Ez2_log}
\end{equation} The derivatives follow analogously.
\paragraph{Derivative with respect to $\Omo$.}
\begin{equation}
\boxed{
\frac{\partial \ln H}{\partial \Omo} = \frac{(1+z)^3 - f_{\rm DE}^{\ln}(z)}{2\,E^2_{\ln}(z)} }.
\label{eq:dlnH_dOm_log}
\end{equation}

\paragraph{Derivatives with respect to $\oo$ and $\oa$.}
Using
\begin{equation}
\frac{\partial f_{\rm DE}^{\ln}}{\partial \oo} = 3\ln(1+z)\,f_{\rm DE}^{\ln}(z), \qquad \frac{\partial f_{\rm DE}^{\ln}}{\partial \oa} = \frac{3}{2}\ln^2(1+z)\,f_{\rm DE}^{\ln}(z),
\label{eq:dfde_log}
\end{equation}
we obtain
\begin{align}
&\boxed{\frac{\partial \ln H}{\partial \oo} = \frac{3(1-\Omo)\,\ln(1+z)\, f_{\rm DE}^{\ln}(z)}{2\,E^2_{\ln}(z)}}, \nonumber \\
&\boxed{\frac{\partial \ln H}{\partial \oa} = \frac{\frac{3}{2}(1-\Omo)\,\ln^2(1+z)\, f_{\rm DE}^{\ln}(z)}{2\,E^2_{\ln}(z)} =
\frac{3(1-\Omo)\,\ln^2(1+z)\, f_{\rm DE}^{\ln}(z)}{4\,E^2_{\ln}(z)}}.
\label{eq:dlnH_dw0dwa_log}
\end{align}

\subsection{Implementation note for figure reproducibility}
For transparency, in our numerical pipeline we evaluate $E^2(z)$ and the derivatives above at the same redshift sampling used in Figs.~\ref{fig2}--\ref{fig5} and compute all plotted sensitivity measures directly from these analytic expressions.


\begin{thebibliography}{99}


\bibitem{SDSS:2005xqv}
D.~J.~Eisenstein \textit{et al.} [SDSS],
``Detection of the Baryon Acoustic Peak in the Large-Scale Correlation Function of SDSS Luminous Red Galaxies,''
Astrophys. J. \textbf{633}, 560-574 (2005)
doi:10.1086/466512
[arXiv:astro-ph/0501171 [astro-ph]].

\bibitem{2dFGRS:2005yhx}
S.~Cole \textit{et al.} [2dFGRS],
``The 2dF Galaxy Redshift Survey: Power-spectrum analysis of the final dataset and cosmological implications,''
Mon. Not. Roy. Astron. Soc. \textbf{362}, 505-534 (2005)
doi:10.1111/j.1365-2966.2005.09318.x
[arXiv:astro-ph/0501174 [astro-ph]].

\bibitem{Blake:2011en}
C.~Blake, E.~Kazin, F.~Beutler, T.~Davis, D.~Parkinson, S.~Brough, M.~Colless, C.~Contreras, W.~Couch and S.~Croom, \textit{et al.}
``The WiggleZ Dark Energy Survey: mapping the distance-redshift relation with baryon acoustic oscillations,''
Mon. Not. Roy. Astron. Soc. \textbf{418}, 1707-1724 (2011)
doi:10.1111/j.1365-2966.2011.19592.x
[arXiv:1108.2635 [astro-ph.CO]].

\bibitem{BOSS:2012dmf}
K.~S.~Dawson \textit{et al.} [BOSS],
``The Baryon Oscillation Spectroscopic Survey of SDSS-III,''
Astron. J. \textbf{145}, 10 (2013)
doi:10.1088/0004-6256/145/1/10
[arXiv:1208.0022 [astro-ph.CO]].

\bibitem{eBOSS:2015jyv}
K.~S.~Dawson \textit{et al.} [eBOSS],
``The SDSS-IV extended Baryon Oscillation Spectroscopic Survey: Overview and Early Data,''
Astron. J. \textbf{151}, 44 (2016)
doi:10.3847/0004-6256/151/2/44
[arXiv:1508.04473 [astro-ph.CO]].

\bibitem{BOSS:2016wmc}
S.~Alam \textit{et al.} [BOSS],
``The clustering of galaxies in the completed SDSS-III Baryon Oscillation Spectroscopic Survey: cosmological analysis of the DR12 galaxy sample,''
Mon. Not. Roy. Astron. Soc. \textbf{470}, no.3, 2617-2652 (2017)
doi:10.1093/mnras/stx721
[arXiv:1607.03155 [astro-ph.CO]].

\bibitem{DES:2024pwq}
T.~M.~C.~Abbott \textit{et al.} [DES],
``Dark Energy Survey: A 2.1\% measurement of the angular baryonic acoustic oscillation scale at redshift zeff=0.85 from the final dataset,''
Phys. Rev. D \textbf{110}, no.6, 063515 (2024)
doi:10.1103/PhysRevD.110.063515
[arXiv:2402.10696 [astro-ph.CO]].

\bibitem{DESI:2024mwx}
A.~G.~Adame \textit{et al.} [DESI],
JCAP \textbf{02}, 021 (2025)
doi:10.1088/1475-7516/2025/02/021
[arXiv:2404.03002 [astro-ph.CO]].

\bibitem{DESI:2025zgx}
M.~Abdul Karim \textit{et al.} [DESI],
``DESI DR2 Results II: Measurements of Baryon Acoustic Oscillations and Cosmological Constraints,''
[arXiv:2503.14738 [astro-ph.CO]].


\bibitem{Planck:2018vyg}
N.~Aghanim \textit{et al.} [Planck],
``Planck 2018 results. VI. Cosmological parameters,''
Astron. Astrophys. \textbf{641}, A6 (2020)
[erratum: Astron. Astrophys. \textbf{652}, C4 (2021)]
doi:10.1051/0004-6361/201833910
[arXiv:1807.06209 [astro-ph.CO]].

\bibitem{Planck:2018jri}
Y.~Akrami \textit{et al.} [Planck],
``Planck 2018 results. X. Constraints on inflation,''
Astron. Astrophys. \textbf{641}, A10 (2020)
doi:10.1051/0004-6361/201833887
[arXiv:1807.06211 [astro-ph.CO]].

\bibitem{Planck:2019kim}
Y.~Akrami \textit{et al.} [Planck],
``Planck 2018 results. IX. Constraints on primordial non-Gaussianity,''
Astron. Astrophys. \textbf{641}, A9 (2020)
doi:10.1051/0004-6361/201935891
[arXiv:1905.05697 [astro-ph.CO]].








\bibitem{Scolnic:2021amr}
D.~Scolnic, D.~Brout, A.~Carr, A.~G.~Riess, T.~M.~Davis, A.~Dwomoh, D.~O.~Jones, N.~Ali, P.~Charvu and R.~Chen, \textit{et al.}
``The Pantheon+ Analysis: The Full Data Set and Light-curve Release,''
Astrophys. J. \textbf{938}, no.2, 113 (2022)
doi:10.3847/1538-4357/ac8b7a
[arXiv:2112.03863 [astro-ph.CO]].

\bibitem{Brout:2022vxf}
D.~Brout, D.~Scolnic, B.~Popovic, A.~G.~Riess, J.~Zuntz, R.~Kessler, A.~Carr, T.~M.~Davis, S.~Hinton and D.~Jones, \textit{et al.}
``The Pantheon+ Analysis: Cosmological Constraints,''
Astrophys. J. \textbf{938}, no.2, 110 (2022)
doi:10.3847/1538-4357/ac8e04
[arXiv:2202.04077 [astro-ph.CO]].

\bibitem{DES:2024jxu}
T.~M.~C.~Abbott \textit{et al.} [DES],
``The Dark Energy Survey: Cosmology Results with \ensuremath{\sim}1500 New High-redshift Type Ia Supernovae Using the Full 5 yr Data Set,''
Astrophys. J. Lett. \textbf{973}, no.1, L14 (2024)
doi:10.3847/2041-8213/ad6f9f
[arXiv:2401.02929 [astro-ph.CO]].



\bibitem{Heymans:2020gsg}
C.~Heymans, T.~Tr\"oster, M.~Asgari, C.~Blake, H.~Hildebrandt, B.~Joachimi, K.~Kuijken, C.~A.~Lin, A.~G.~S\'anchez and J.~L.~van den Busch, \textit{et al.}
``KiDS-1000 Cosmology: Multi-probe weak gravitational lensing and spectroscopic galaxy clustering constraints,''
Astron. Astrophys. \textbf{646}, A140 (2021)
doi:10.1051/0004-6361/202039063
[arXiv:2007.15632 [astro-ph.CO]].

\bibitem{DES:2021bvc}
A.~Amon \textit{et al.} [DES],
``Dark Energy Survey Year 3 results: Cosmology from cosmic shear and robustness to data calibration,''
Phys. Rev. D \textbf{105}, no.2, 023514 (2022)
doi:10.1103/PhysRevD.105.023514
[arXiv:2105.13543 [astro-ph.CO]].

\bibitem{Dalal:2023olq}
R.~Dalal, X.~Li, A.~Nicola, J.~Zuntz, M.~A.~Strauss, S.~Sugiyama, T.~Zhang, M.~M.~Rau, R.~Mandelbaum and M.~Takada, \textit{et al.}
``Hyper Suprime-Cam Year 3 results: Cosmology from cosmic shear power spectra,''
Phys. Rev. D \textbf{108}, no.12, 123519 (2023)
doi:10.1103/PhysRevD.108.123519
[arXiv:2304.00701 [astro-ph.CO]].

\bibitem{Sugiyama:2023fzm}
S.~Sugiyama, H.~Miyatake, S.~More, X.~Li, M.~Shirasaki, M.~Takada, Y.~Kobayashi, R.~Takahashi, T.~Nishimichi and A.~J.~Nishizawa, \textit{et al.}
``Hyper Suprime-Cam Year 3 results: Cosmology from galaxy clustering and weak lensing with HSC and SDSS using the minimal bias model,''
Phys. Rev. D \textbf{108}, no.12, 123521 (2023)
doi:10.1103/PhysRevD.108.123521
[arXiv:2304.00705 [astro-ph.CO]].


\bibitem{Bhattacharya:2010cf}
S.~Bhattacharya, A.~Kosowsky, J.~A.~Newman and A.~R.~Zentner,
``Galaxy Peculiar Velocities From Large-Scale Supernova Surveys as a Dark Energy Probe,''
Phys. Rev. D \textbf{83}, 043004 (2011)
doi:10.1103/PhysRevD.83.043004
[arXiv:1008.2560 [astro-ph.CO]].

\bibitem{Agrawal:2019yed}
A.~Agrawal, T.~Okumura and T.~Futamase,
``Constraining neutrino mass and dark energy with peculiar velocities and lensing dispersions of Type Ia supernovae,''
Phys. Rev. D \textbf{100}, no.6, 063534 (2019)
doi:10.1103/PhysRevD.100.063534
[arXiv:1907.02328 [astro-ph.CO]].

\bibitem{Zhang:2024pyf}
W.~Zhang, M.~c.~Chu, S.~Liao, S.~Yeung and H.~J.~Hu,
``Measuring the Hubble Constant through the Galaxy Pairwise Peculiar Velocity,''
Astrophys. J. Lett. \textbf{978}, no.1, L6 (2025)
doi:10.3847/2041-8213/ad9aa7
[arXiv:2412.04660 [astro-ph.CO]].



\bibitem{Chevallier:2000qy}
M.~Chevallier and D.~Polarski,
``Accelerating universes with scaling dark matter,''
Int. J. Mod. Phys. D \textbf{10}, 213-224 (2001)
doi:10.1142/S0218271801000822
[arXiv:gr-qc/0009008 [gr-qc]].

\bibitem{Linder:2002et}
E.~V.~Linder,
``Exploring the expansion history of the universe,''
Phys. Rev. Lett. \textbf{90}, 091301 (2003)
doi:10.1103/PhysRevLett.90.091301
[arXiv:astro-ph/0208512 [astro-ph]].


\bibitem{Myrzakulov:2023qjo}
N.~Myrzakulov, M.~Koussour and D.~J.~Gogoi,
Phys. Dark Univ. \textbf{42}, 101268 (2023)
doi:10.1016/j.dark.2023.101268
[arXiv:2306.13218 [gr-qc]].

\bibitem{Myrzakulov:2023sir}
N.~Myrzakulov, M.~Koussour and A.~Mussatayeva,
Chin. J. Phys. \textbf{85}, 345-358 (2023)
doi:10.1016/j.cjph.2023.07.003
[arXiv:2308.15101 [gr-qc]].

\bibitem{Shekh:2023vtq}
S.~H.~Shekh, N.~Myrzakulov, A.~Pradhan and A.~Mussatayeva,
Symmetry \textbf{15}, no.2, 321 (2023)
doi:10.3390/sym15020321









\bibitem{Seo:2003pu}
H.~J.~Seo and D.~J.~Eisenstein,
``Probing dark energy with baryonic acoustic oscillations from future large galaxy redshift surveys,''
Astrophys. J. \textbf{598}, 720-740 (2003)
doi:10.1086/379122
[arXiv:astro-ph/0307460 [astro-ph]].

\bibitem{Gaztanaga:2008xz}
E.~Gaztanaga, A.~Cabre and L.~Hui,
``Clustering of Luminous Red Galaxies IV: Baryon Acoustic Peak in the Line-of-Sight Direction and a Direct Measurement of H(z),''
Mon. Not. Roy. Astron. Soc. \textbf{399}, 1663-1680 (2009)
doi:10.1111/j.1365-2966.2009.15405.x
[arXiv:0807.3551 [astro-ph]].

\bibitem{Moresco:2012by}
M.~Moresco, L.~Verde, L.~Pozzetti, R.~Jimenez and A.~Cimatti,
``New constraints on cosmological parameters and neutrino properties using the expansion rate of the Universe to z\textasciitilde{}1.75,''
JCAP \textbf{07}, 053 (2012)
doi:10.1088/1475-7516/2012/07/053
[arXiv:1201.6658 [astro-ph.CO]].







\bibitem{Jimenez:2003iv}
R.~Jimenez, L.~Verde, T.~Treu and D.~Stern,
``Constraints on the equation of state of dark energy and the Hubble constant from stellar ages and the CMB,''
Astrophys. J. \textbf{593}, 622-629 (2003)
doi:10.1086/376595
[arXiv:astro-ph/0302560 [astro-ph]].

\bibitem{Simon:2004tf}
J.~Simon, L.~Verde and R.~Jimenez,
``Constraints on the redshift dependence of the dark energy potential,''
Phys. Rev. D \textbf{71}, 123001 (2005)
doi:10.1103/PhysRevD.71.123001
[arXiv:astro-ph/0412269 [astro-ph]].

\bibitem{Stern:2009ep}
D.~Stern, R.~Jimenez, L.~Verde, M.~Kamionkowski and S.~A.~Stanford,
``Cosmic Chronometers: Constraining the Equation of State of Dark Energy. I: H(z) Measurements,''
JCAP \textbf{02}, 008 (2010)
doi:10.1088/1475-7516/2010/02/008
[arXiv:0907.3149 [astro-ph.CO]].

\bibitem{Borghi:2021rft}
N.~Borghi, M.~Moresco and A.~Cimatti,
Astrophys. J. Lett. \textbf{928}, no.1, L4 (2022)
doi:10.3847/2041-8213/ac3fb2
[arXiv:2110.04304 [astro-ph.CO]].

\bibitem{Jiao:2022aep}
K.~Jiao, N.~Borghi, M.~Moresco and T.~J.~Zhang,
Astrophys. J. Suppl. \textbf{265}, no.2, 48 (2023)
doi:10.3847/1538-4365/acbc77
[arXiv:2205.05701 [astro-ph.CO]].

\bibitem{Moresco:2024wmr}
M.~Moresco,
``Measuring the expansion history of the Universe with cosmic chronometers,''
[arXiv:2412.01994 [astro-ph.CO]].


\bibitem{Ma:2010mr}
C.~Ma and T.~J.~Zhang,
``Power of Observational Hubble Parameter Data: a Figure of Merit Exploration,''
Astrophys. J. \textbf{730}, 74 (2011)
doi:10.1088/0004-637X/730/2/74
[arXiv:1007.3787 [astro-ph.CO]].


\bibitem{Cooray:2005yr}
A.~Cooray, D.~Huterer and D.~Holz,
``Problems with pencils: lensing covariance of supernova distance measurements,''
Phys. Rev. Lett. \textbf{96}, 021301 (2006)
doi:10.1103/PhysRevLett.96.021301
[arXiv:astro-ph/0509581 [astro-ph]].



















\bibitem{Wang:2005yaa}
Y.~Wang and M.~Tegmark,
Phys. Rev. D \textbf{71}, 103513 (2005)
doi:10.1103/PhysRevD.71.103513
[arXiv:astro-ph/0501351 [astro-ph]].





\bibitem{Hernandez-Monteagudo:2005xtx}
C.~Hernandez-Monteagudo, L.~Verde, R.~Jimenez and D.~N.~Spergel,
``Correlation properties of the kinematic sunyaev-zel'dovich effect and implications for dark energy,''
Astrophys. J. \textbf{643}, 598-615 (2006)
doi:10.1086/503190
[arXiv:astro-ph/0511061 [astro-ph]].

\bibitem{Bhattacharya:2007sk}
S.~Bhattacharya and A.~Kosowsky,
``Dark Energy Constraints from Galaxy Cluster Peculiar Velocities,''
Phys. Rev. D \textbf{77}, 083004 (2008)
doi:10.1103/PhysRevD.77.083004
[arXiv:0712.0034 [astro-ph]].

\bibitem{Hand:2012ui}
N.~Hand, G.~E.~Addison, E.~Aubourg, N.~Battaglia, E.~S.~Battistelli, D.~Bizyaev, J.~R.~Bond, H.~Brewington, J.~Brinkmann and B.~R.~Brown, \textit{et al.}
``Evidence of Galaxy Cluster Motions with the Kinematic Sunyaev-Zel'dovich Effect,''
Phys. Rev. Lett. \textbf{109}, 041101 (2012)
doi:10.1103/PhysRevLett.109.041101
[arXiv:1203.4219 [astro-ph.CO]].

\bibitem{Tanimura:2020une}
H.~Tanimura, S.~Zaroubi and N.~Aghanim,
``Direct detection of the kinetic Sunyaev-Zel'dovich effect in galaxy clusters,''
Astron. Astrophys. \textbf{645}, A112 (2021)
doi:10.1051/0004-6361/202038846
[arXiv:2007.02952 [astro-ph.CO]].


\bibitem{DES:2023mug}
M.~Mallaby-Kay \textit{et al.} [DES],
``Kinematic Sunyaev-Zel\textquoteright{}dovich effect with ACT, DES, and BOSS: A novel hybrid estimator,''
Phys. Rev. D \textbf{108}, no.2, 023516 (2023)
doi:10.1103/PhysRevD.108.023516
[arXiv:2305.06792 [astro-ph.CO]].


\bibitem{Lavaux:2007zw}
G.~Lavaux, R.~Mohayaee, S.~Colombi, R.~B.~Tully, F.~Bernardeau and J.~Silk,
Mon. Not. Roy. Astron. Soc. \textbf{383}, 1292 (2008)
doi:10.1111/j.1365-2966.2007.12539.x
[arXiv:0707.3483 [astro-ph]].


\bibitem{Alfedeel:2024ktc}
A.~H.~A.~Alfedeel, M.~Koussour and N.~Myrzakulov,
Astron. Comput. \textbf{47}, 100821 (2024)
doi:10.1016/j.ascom.2024.100821
[arXiv:2403.09916 [gr-qc]].

\bibitem{Koussour:2024kxd}
M.~Koussour, N.~Myrzakulov and M.~K.~M.~Ali,
JHEAp \textbf{42}, 96-103 (2024)
doi:10.1016/j.jheap.2024.04.003
[arXiv:2404.03362 [astro-ph.CO]].












\bibitem{SDSS:2009ocz}
W.~J.~Percival \textit{et al.} [SDSS],
Mon. Not. Roy. Astron. Soc. \textbf{401}, 2148-2168 (2010)
doi:10.1111/j.1365-2966.2009.15812.x
[arXiv:0907.1660 [astro-ph.CO]].

\bibitem{BOSS:2013rlg}
L.~Anderson \textit{et al.} [BOSS],
Mon. Not. Roy. Astron. Soc. \textbf{441}, no.1, 24-62 (2014)
doi:10.1093/mnras/stu523
[arXiv:1312.4877 [astro-ph.CO]].









\bibitem{Chantry:2010ym}
V.~Chantry, D.~Sluse and P.~Magain,
``COSMOGRAIL: the COSmological MOnitoring of GRAvItational Lenses VIII. Deconvolution of high resolution near-IR images and simple mass models for 7 gravitationally lensed quasars,''
Astron. Astrophys. \textbf{522}, A95 (2010)
doi:10.1051/0004-6361/200912971
[arXiv:1007.3142 [astro-ph.CO]].


\bibitem{H0LiCOW:2019pvv}
K.~C.~Wong \textit{et al.} [H0LiCOW],
``H0LiCOW \textendash{} XIII. A 2.4 per cent measurement of H0 from lensed quasars: 5.3\ensuremath{\sigma} tension between early- and late-Universe probes,''
Mon. Not. Roy. Astron. Soc. \textbf{498}, no.1, 1420-1439 (2020)
doi:10.1093/mnras/stz3094
[arXiv:1907.04869 [astro-ph.CO]].












\bibitem{SupernovaSearchTeam:1998fmf}
A.~G.~Riess \textit{et al.} [Supernova Search Team],
``Observational evidence from supernovae for an accelerating universe and a cosmological constant,''
Astron. J. \textbf{116}, 1009-1038 (1998)
doi:10.1086/300499
[arXiv:astro-ph/9805201 [astro-ph]].

\bibitem{SupernovaCosmologyProject:1998vns}
S.~Perlmutter \textit{et al.} [Supernova Cosmology Project],
``Measurements of $\Omega$ and $\Lambda$ from 42 High Redshift Supernovae,''
Astrophys. J. \textbf{517}, 565-586 (1999)
doi:10.1086/307221
[arXiv:astro-ph/9812133 [astro-ph]].

\bibitem{Pan-STARRS1:2017jku}
D.~M.~Scolnic \textit{et al.} [Pan-STARRS1],
``The Complete Light-curve Sample of Spectroscopically Confirmed SNe Ia from Pan-STARRS1 and Cosmological Constraints from the Combined Pantheon Sample,''
Astrophys. J. \textbf{859}, no.2, 101 (2018)
doi:10.3847/1538-4357/aab9bb
[arXiv:1710.00845 [astro-ph.CO]].
  











\bibitem{Zhang:2007nk}
P.~Zhang, M.~Liguori, R.~Bean and S.~Dodelson,
``Probing Gravity at Cosmological Scales by Measurements which Test the Relationship between Gravitational Lensing and Matter Overdensity,''
Phys. Rev. Lett. \textbf{99}, 141302 (2007)
doi:10.1103/PhysRevLett.99.141302
[arXiv:0704.1932 [astro-ph]].

\bibitem{Reyes:2010tr}
R.~Reyes, R.~Mandelbaum, U.~Seljak, T.~Baldauf, J.~E.~Gunn, L.~Lombriser and R.~E.~Smith,
``Confirmation of general relativity on large scales from weak lensing and galaxy velocities,''
Nature \textbf{464}, 256-258 (2010)
doi:10.1038/nature08857
[arXiv:1003.2185 [astro-ph.CO]].

\bibitem{Lee:2025kbn}
S.~Lee,
Mon. Not. Roy. Astron. Soc. \textbf{544}, 3388-3393 (2025)
doi:10.1093/mnras/staf1890
[arXiv:2506.16022 [astro-ph.CO]].


\bibitem{Lee:2011ec}
S.~Lee,
Astrophys. Space Sci. \textbf{350}, 785-790 (2014)
doi:10.1007/s10509-013-1762-1
[arXiv:1105.0993 [astro-ph.CO]].

\bibitem{Colgain:2021pmf}
E.~\'O.~Colg\'ain, M.~M.~Sheikh-Jabbari and L.~Yin,
Phys. Rev. D \textbf{104}, no.2, 023510 (2021)
doi:10.1103/PhysRevD.104.023510
[arXiv:2104.01930 [astro-ph.CO]].



\bibitem{Efstathiou:1999tm}
G.~Efstathiou,
Mon. Not. Roy. Astron. Soc. \textbf{310}, 842-850 (1999)
doi:10.1046/j.1365-8711.1999.02997.x
[arXiv:astro-ph/9904356 [astro-ph]].


\bibitem{Lee:2010nb}
S.~Lee,
[arXiv:1005.1770 [astro-ph.CO]].




\bibitem{Astier:2000as}
P.~Astier,
``Can luminosity distance measurements probe the equation of state of dark energy,''
Phys. Lett. B \textbf{500}, 8-15 (2001)
doi:10.1016/S0370-2693(01)00072-7
[arXiv:astro-ph/0008306 [astro-ph]].


\bibitem{Huterer:2000mj}
D.~Huterer and M.~S.~Turner,
``Probing the dark energy: Methods and strategies,''
Phys. Rev. D \textbf{64}, 123527 (2001)
doi:10.1103/PhysRevD.64.123527
[arXiv:astro-ph/0012510 [astro-ph]].

\bibitem{Maor:2001ku}
I.~Maor, R.~Brustein, J.~McMahon and P.~J.~Steinhardt,
``Measuring the equation of state of the universe: Pitfalls and prospects,''
Phys. Rev. D \textbf{65}, 123003 (2002)
doi:10.1103/PhysRevD.65.123003
[arXiv:astro-ph/0112526 [astro-ph]].

\bibitem{Colgain:2022tql}
E.~\'O.~Colg\'ain, M.~M.~Sheikh-Jabbari and R.~Solomon,
Phys. Dark Univ. \textbf{40}, 101216 (2023)
doi:10.1016/j.dark.2023.101216
[arXiv:2211.02129 [astro-ph.CO]].





\bibitem{Alam:2003fg}
U.~Alam, V.~Sahni, T.~D.~Saini and A.~A.~Starobinsky,
``Is there supernova evidence for dark energy metamorphosis ?,''
Mon. Not. Roy. Astron. Soc. \textbf{354}, 275 (2004)
doi:10.1111/j.1365-2966.2004.08189.x
[arXiv:astro-ph/0311364 [astro-ph]].


\bibitem{Wu:2007tz}
P.~Wu and H.~W.~Yu,
``Reconstructing the properties of dark energy from recent observations,''
JCAP \textbf{10}, 014 (2007)
doi:10.1088/1475-7516/2007/10/014
[arXiv:0710.1958 [astro-ph]].








\bibitem{Samushia:2006fx}
L.~Samushia and B.~Ratra,
Astrophys. J. Lett. \textbf{650}, L5-L8 (2006)
doi:10.1086/508662
[arXiv:astro-ph/0607301 [astro-ph]].








\bibitem{Lee:2009gb}
S.~Lee and K.~W.~Ng,
``Growth index with the exact analytic solution of sub-horizon scale linear perturbation for dark energy models with constant equation of state,''
Phys. Lett. B \textbf{688}, 1-3 (2010)
doi:10.1016/j.physletb.2010.03.082
[arXiv:0906.1643 [astro-ph.CO]].


\bibitem{Lee:2009ft}
S.~Lee and K.~W.~Ng,
``Properties of the exact analytic solution of the growth factor and its applications,''
Phys. Rev. D \textbf{82}, 043004 (2010)
doi:10.1103/PhysRevD.82.043004
[arXiv:0907.2108 [astro-ph.CO]].

\bibitem{Lee:2009if}
S.~Lee and K.~W.~Ng,
Chin. J. Phys. \textbf{50}, 367 (2012)
[arXiv:0905.1522 [astro-ph.CO]].



\bibitem{Colgain:2025nzf}
E.~\'O.~Colg\'ain, S.~Pourojaghi, M.~M.~Sheikh-Jabbari and L.~Yin,
[arXiv:2504.04417 [astro-ph.CO]].


\bibitem{Pedrotti:2024kpn}
D.~Pedrotti, J.~Q.~Jiang, L.~A.~Escamilla, S.~S.~da Costa and S.~Vagnozzi,
Phys. Rev. D \textbf{111}, no.2, 023506 (2025)
doi:10.1103/PhysRevD.111.023506
[arXiv:2408.04530 [astro-ph.CO]].

\bibitem{Wang:2025vfb}
Y.~Wang and K.~Freese,
[arXiv:2505.17415 [astro-ph.CO]].

\end{thebibliography}
\end{document}